\documentclass[aps,prl,reprint,superscriptaddress,showpacs,longbibliography,footnoteinbib]{revtex4-1}
\usepackage[latin9]{inputenc}
\usepackage{color}
\usepackage{bm}
\usepackage{amsmath}
\usepackage{amssymb}
\usepackage{graphicx}
\usepackage[unicode=true,
 bookmarks=false,
 breaklinks=false,pdfborder={0 0 1},backref=false,colorlinks=true]
 {hyperref}
\hypersetup{pdftitle={Title},
 plainpages=false,pdfpagelabels,linkcolor=blue,urlcolor=blue,citecolor=blue,pdfdisplaydoctitle=true,pdfduplex=DuplexFlipLongEdge}

\makeatletter
\usepackage{units}\usepackage{wasysym}

\definecolor{orange}{rgb}{0.50, 0.20, 0.0}

\newcommand{\beginsupplement}{%
	\setcounter{page}{1}
	 \renewcommand{\thepage}{SM - \arabic{page}}%
        \setcounter{table}{0}
        \renewcommand{\thetable}{S\arabic{table}}%
        \setcounter{figure}{0}
        \renewcommand{\thefigure}{S\arabic{figure}}%
        \setcounter{section}{0}
        \renewcommand{\thesection}{S\arabic{section}}%
        \setcounter{section}{0}
        \renewcommand{\thesection}{S\arabic{section}}%
        \setcounter{subsection}{0}
        \renewcommand{\thesubsection}{S\arabic{section}.\arabic{subsection}}%
        \setcounter{equation}{0}
        \renewcommand{\theequation}{S\arabic{equation}}%

     }

\usepackage{bm}
\usepackage{braket}

\makeatother

\begin{document}
\noindent\begin{minipage}[t]{1\columnwidth}%
\global\long\def\ket#1{\left| #1\right\rangle }%

\global\long\def\bra#1{\left\langle #1 \right|}%

\global\long\def\kket#1{\left\Vert #1\right\rangle }%

\global\long\def\bbra#1{\left\langle #1\right\Vert }%

\global\long\def\braket#1#2{\left\langle #1\right. \left| #2 \right\rangle }%

\global\long\def\bbrakket#1#2{\left\langle #1\right. \left\Vert #2\right\rangle }%

\global\long\def\av#1{\left\langle #1 \right\rangle }%

\global\long\def\tr{\text{tr}}%

\global\long\def\Tr{\text{Tr}}%

\global\long\def\pd{\partial}%

\global\long\def\im{\text{Im}}%

\global\long\def\re{\text{Re}}%

\global\long\def\sgn{\text{sgn}}%

\global\long\def\Det{\text{Det}}%

\global\long\def\abs#1{\left|#1\right|}%

\global\long\def\up{\uparrow}%

\global\long\def\down{\downarrow}%

\global\long\def\vc#1{\mathbf{#1}}%

\global\long\def\bs#1{\boldsymbol{#1}}%

\global\long\def\t#1{\text{#1}}%
\end{minipage}

\title{Critical phase dualities in 1D exactly-solvable quasiperiodic models}
\author{Miguel Gonçalves}
\affiliation{CeFEMA, LaPMET, Instituto Superior Técnico, Universidade de Lisboa,
Av. Rovisco Pais, 1049-001 Lisboa, Portugal}
\author{Bruno Amorim}
\affiliation{Centro de Física das Universidades do Minho e Porto, LaPMET, University
of Minho, Campus of Gualtar, 4710-057, Braga, Portugal}
\author{Eduardo V. Castro}
\affiliation{Centro de Física das Universidades do Minho e Porto, LaPMET, Departamento
de Física e Astronomia, Faculdade de Ciências, Universidade do Porto,
4169-007 Porto, Portugal}
\affiliation{Beijing Computational Science Research Center, Beijing 100193, China}
\author{Pedro Ribeiro}
\affiliation{CeFEMA, LaPMET, Instituto Superior Técnico, Universidade de Lisboa,
Av. Rovisco Pais, 1049-001 Lisboa, Portugal}
\affiliation{Beijing Computational Science Research Center, Beijing 100193, China}
\begin{abstract}
We propose a solvable class of 1D quasiperiodic tight-binding models
encompassing extended, localized, and critical phases, separated by
nontrivial mobility edges. Limiting cases include the Aubry-André
model and the models of PRL 114, 146601 and PRL 104, 070601. The analytical
treatment follows from recognizing these models as a novel type of
fixed-points of the renormalization group procedure recently proposed
in arXiv:2206.13549  for characterizing phases of quasiperiodic structures.
Beyond known limits, the proposed class of models extends previously
encountered localized-delocalized duality transformations to points
within multifractal critical phases. Besides an experimental confirmation
of multifractal duality, realizing the proposed class of models in
optical lattices allows stabilizing multifractal critical phases and
non-trivial mobility edges without the need for the unbounded potentials
required by previous proposals. 
\end{abstract}
\maketitle
Quasiperiodic systems (QPS) offer a rich playground of interesting
physics ranging from exotic localization properties in one \citep{AubryAndre,Roati2008,Lahini2009,Schreiber842,Luschen2018,slager1}
or higher \citep{Huang2016a,PhysRevLett.120.207604,Park2018,PhysRevB.100.144202,Fu2020,Wang2020,goncalves2020,PhysRevX.7.041047}
dimensions, to intriguing topological properties \citep{Kraus2012,PhysRevLett.109.116404,Verbin2013,Zilberberg:21,slager2}.
Quasiperiodicity has been studied in widely different platforms, including
optical \citep{PhysRevA.75.063404,Roati2008,Modugno_2009,Schreiber842,Luschen2018,PhysRevLett.123.070405,PhysRevLett.125.060401,PhysRevLett.126.110401,PhysRevLett.126.040603,PhysRevLett.122.170403}
and photonic lattices \citep{Lahini2009,Kraus2012,Verbin2013,PhysRevB.91.064201,Wang2020,https://doi.org/10.1002/adom.202001170,Wang2022},
cavity-polariton devices \citep{Goblot2020Nature}, phononic media
\citep{PhysRevLett.122.095501,Ni2019,PhysRevLett.125.224301,PhysRevApplied.13.014023,PhysRevX.11.011016,doi:10.1063/5.0013528},
moiré materials \citep{Balents2020}, periodically and quasiperiodically-driven
systems \citep{PhysRevB.96.144301,10.21468/SciPostPhys.4.5.025,CadeZ2019,Bordia2017,PhysRevB.103.184309,Dumitrescu2022,kicked_QC_2022},
and non-hermitian quasicrystals \citep{PhysRevB.100.054301,PhysRevLett.122.237601,PhysRevX.9.041015,PhysRevB.103.014203,PhysRevB.104.024201,PhysRevB.103.134208,PhysRevLett.129.113601}.
The ubiquity of QPS and their relevance to several interdisciplinary
topical issues rendered these systems a hot topic of research.

QPS host phases with fully localized and extended wave-functions.
Interestingly, quasiperiodicity can also stabilize critical multifractal
states, first encountered at the localization-delocatization transition
lines, and later found to persist over extended regions \citep{PhysRevLett.110.146404,Liu2015,PhysRevB.93.104504,CadeZ2019,PhysRevLett.123.025301,PhysRevLett.125.073204,anomScipost,PhysRevB.106.024204}.

QPS present substantial challenges for theoretical methods, and an
analytical treatment of the localization phase diagrams remains restricted
to a few fine-tuned models \citep{AubryAndre,PhysRevB.43.13468,PhysRevLett.104.070601,PhysRevLett.113.236403,Liu2015,PhysRevB.91.235134,PhysRevLett.114.146601},
and even a smaller subset hosts critical phases \citep{Liu2015,anomScipost}.
In particular, Ref.~\citep{Liu2015} found critical phases with energy-independent
transitions to localized and delocalized phases, i.e. without mobility
edges. These were shown to be robust to interactions, giving rise
to many-body critical regimes \citep{PhysRevLett.126.080602} and
have been simulated using ultracold atoms \citep{XIAO20212175}. In
Ref.~\citep{anomScipost} mobility edges were reported, however,
requiring unbounded potentials. Examples of co-existence of extended,
critical and localized regimes, separated by mobility edges, were
reported in Ref. \citep{mobedges_ext_crit_loc}, but only numerically.
As the existence of energy-dependent critical-to-extended or critical-to-localized
transitions has not been experimentally reported so far, more models
with such physics, no need for diverging potentials, and with analytically
exact phase diagrams, are of topical and practical interest for experimental
implementations.

Here, we propose a class of 1D quasiperiodic tight binding models
that includes extended, localized and critical phases and determine
its phase diagram analytically in the thermodynamic limit. Physically
motivated by previous experimentally realizations in optical lattices,
our models contain exponentially decaying hoppings and quasiperiodic
harmonics, with a tunable decay length. As limiting cases, this class
contains the Aubry-André model and the model in Ref.~\citep{PhysRevLett.114.146601},
that were already experimentally realized \citep{Roati2008,PhysRevLett.126.040603},
and the model in Ref.$\,$\citep{PhysRevLett.104.070601}. Away from
these limits, our class of models contains novel features, not found
in any of the limiting cases: critical phases that extend over a considerable
region of parameters and energy-dependent transitions between critical
and extended or localized phases.

\begin{figure}[h]
\begin{centering}
\includegraphics[width=1\columnwidth]{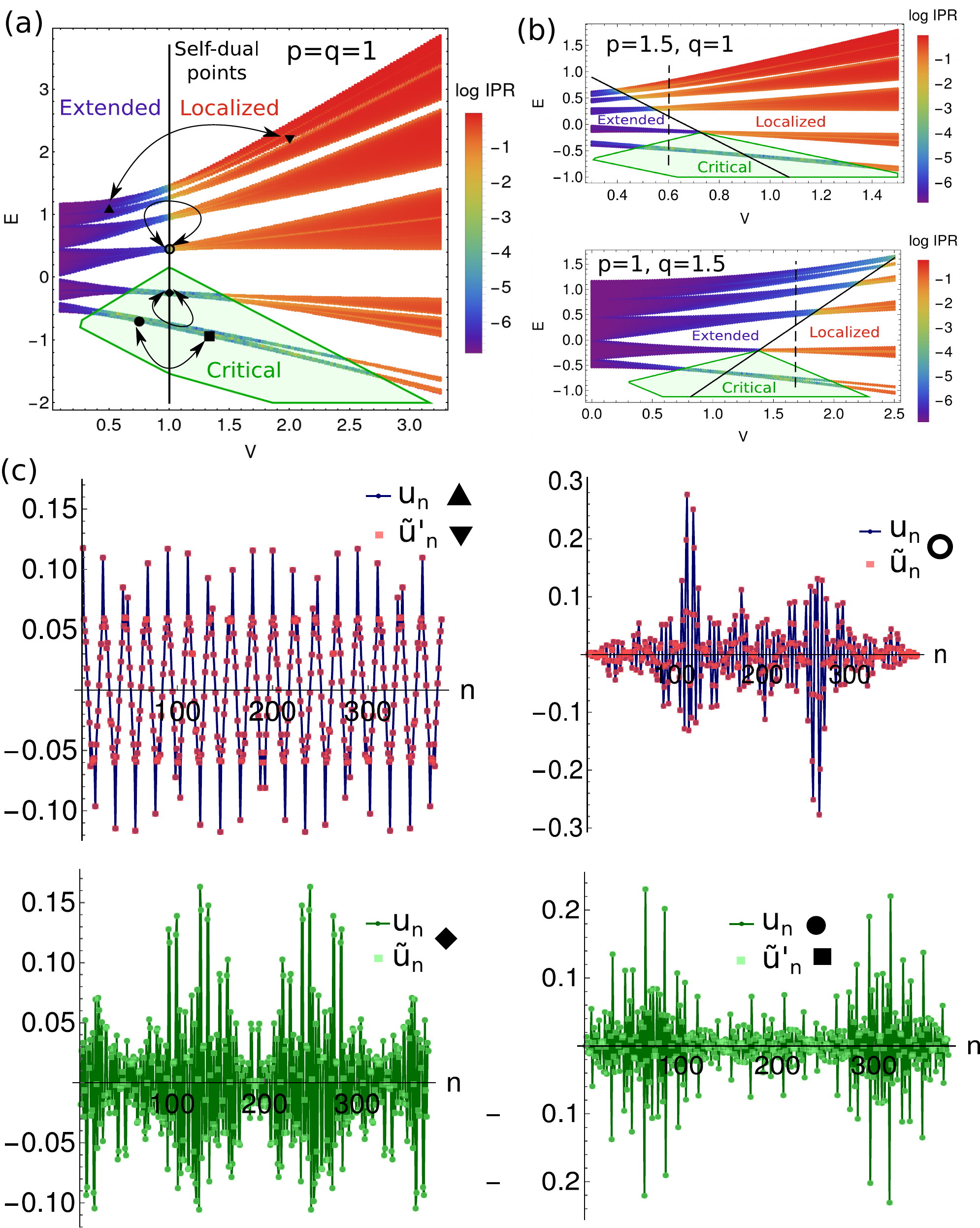}
\par\end{centering}
\caption{(a) IPR (see below Eq.$\,$\eqref{eq:model-k} for definition) results
obtained numerically for $L=F_{16}=987$, $p=q=1$ and as a function
of the strength of the quasiperiodic potential $V$ (see Eq.$\,$\ref{eq:model}).
Superimposed are the analytical extended-localized phase boundaries
(SD points) and the critical phase (bounded by green lines). (b) Phase
diagrams obtained for $p=1.5,q=1$ (up) and $p=1,q=1.5$ (down). The
dashed lines indicate values of $V$ for which all the phases can
be reached at different energies. (c) Examples of eigenstates $u_{n}$
and dual eigenstates $\tilde{u}_{n}$ defined in Eq.$\,$\eqref{eq:duality_transformation}
at dual points in the phase diagram indicated in (a), for $L=F_{14}=377$.
Since $p=q$, $W(x)=1$ and $\tilde{u}'_{n}$ and $u_{n}$ are simply
related by the Aubry-André duality. \label{fig:dualities}}
\end{figure}

The main results are shown in Fig.$\,$\ref{fig:dualities}. In Fig.$\,$\ref{fig:dualities}(a),
we show numerical and analytical results for the phase diagram, for
a fixed set of parameters, where a critical phase exists over a wide
range of the quasiperiodic potential strength $V$ (see Eq.$\,$\ref{eq:model}).
The phase diagram hosts exact dualities that are more general than
the ones previously found for the limiting models in Refs.~\citep{PhysRevLett.104.070601,PhysRevLett.114.146601}.
They exist not only between the extended and localized phase and at
the self-dual (SD) transition points, but also within the critical
phase. Examples are shown in Fig.$\,$\ref{fig:dualities}(c), where
the real-space wave function amplitude $u_{n}$ at site $n$, is exactly
equal to its dual $\tilde{u}_{n}$ (see Eq.$\,$\ref{eq:duality_transformation}
for definition), at dual points in the phase diagram marked in Fig.$\,$\ref{fig:dualities}(a).
In Fig.$\,$\ref{fig:dualities}(b), we also show that highly tunable
mobility edges between extended and localized phases can be introduced
by choosing different decay lengths for the hoppings and quasiperiodic
harmonics. Interestingly, we can have all the phases, including the
critical phase, arising at different energies for a fixed set of parameters,
as also shown in Fig.$\,$\ref{fig:dualities}(b).

\paragraph*{Model and methods.---}

We consider a family of models parameterized by the Hamiltonian

\begin{equation}
\begin{aligned}H= & t\sum_{n\neq n'}e^{i\alpha(n-n')}e^{-p|n-n'|}c_{n}^{\dagger}c_{n'}\\
 & +2V\sum_{n}\sum_{l=1}^{+\infty}e^{-ql}\cos[l(2\pi\tau n+\phi)]c_{n}^{\dagger}c_{n}
\end{aligned}
\label{eq:model}
\end{equation}
where $c_{n}^{\dagger}$ creates a particle at site $n$. The first
term describes hoppings modulated by a magnetic flux $\alpha$ with
an exponential decay determined by $p$. The second term represents
a quasiperiodic potential, incommensurate with the lattice for $\tau\neq\mathbb{Q}$,
obtained by summing harmonics of the incommensurate wavenumber $2\pi\tau$
with exponentially decaying amplitudes controlled by the parameter
$q$. In the following we set $t=1$ unless otherwise stated. The
model in Eq.$\,$\eqref{eq:model} reduces to that in \citep{PhysRevLett.104.070601}
in the $q\to\infty$ limit and $\alpha=0$ after the replacing $t\rightarrow te^{p}$
and $V\rightarrow Ve^{q}$. Similarly, it reduces to the model in
\citep{PhysRevLett.114.146601} for large $p$, and to the Aubry-André
model when both $p$ and $q$ are large.

We consider finite systems with $L$ sites. In order to avoid boundary
defects, we consider rational approximants of the irrational parameter
$\tau$. We chose $1/\tau$ as the golden ratio in the numerical calculations,
but our analytical results for the phase diagram are independent of
$\tau$. The rational approximants are written as $\tau_{c}^{(n)}=F_{n-1}/F_{n}$,
where $F_{n}$ is a Fibonacci number defining the number of sites
$L$ in the unit cell, with $L=F_{n}$ \citep{PhysRevLett.43.1954,PhysRevLett.51.1198}.
We impose twisted boundary conditions, with phase twists $k$ which
is the same as working in a fixed momentum sector of the Hamiltonian
in the Bloch basis defined as $c_{n}\rightarrow c_{m,r}=N^{-1/2}\sum_{k}e^{ik(m+rL)}\tilde{c}_{m,k}$,
where $m=0,\cdots,L-1$ runs over the $L$ sites of the unit cell,
and $r=0,\cdots,N$ is the unit cell index, with $N\rightarrow\infty$
the total number of unit cells. The Hamiltonian for a fixed $k$-sector
becomes

{\footnotesize{}
\begin{equation}
\begin{aligned}H(k)= & t\sum_{r=-\infty}^{\infty}\sum_{m,m'=0}^{L-1}e^{-p|rL+m-m'|}e^{i(\alpha-k)(m+rL-m')}\tilde{c}_{m,k}^{\dagger}\tilde{c}_{m',k}\\
 & +2V\sum_{m=0}^{L-1}\sum_{l=1}^{+\infty}e^{-ql}\cos[l(2\pi\tau_{c}m+\phi)]\tilde{c}_{m,k}^{\dagger}\tilde{c}_{m,k}
\end{aligned}
.\label{eq:model-k}
\end{equation}
}which is just the Hamiltonian of a system with $L$ sites and a phase
twist $k$. For the analytical calculations, we study commensurate
approximants (CA) defined by $\tau_{c}=L'/L$, where $L'$ and $L$
are co-prime integers, in the $L\rightarrow\infty$ limit (infinite
unit cell size/quasiperiodic limit). In particular, we use the methods
introduced in Ref.~\citep{GoncalvesRG2022} and an exact generalized
duality that we prove below.

Our analytical results are confirmed numerically through the real-space
and momentum-space inverse participation ratios, respectively ${\rm IPR}$
and ${\rm IPR}_{k}$. For an eigenstate $\ket{\psi(E)}=\sum_{n}\psi_{n}(E)\ket n$,
where $\{\ket n\}$ is a basis localized at each site, these quantities
are defined as ${\rm IPR}_{(k)}(E)=(\sum_{n}|\psi_{n}^{(k)}(E)|^{2})^{-2}\sum_{n}|\psi_{n}^{(k)}(E)|^{4}$
\citep{Aulbach_2004}, where $\psi_{n}^{k}(E)$ are the amplitudes
of the discrete Fourier transform of the set $\left\{ \psi_{n}(E)\right\} $.
In the extended phase, the ${\rm IPR}$ scales as $L^{-1}$ and ${\rm IPR}_{k}$
is $L$-independent, while in the localized phase, the ${\rm IPR}_{k}$
scales as $L^{-1}$ while the ${\rm IPR}$ is $L$-independent (for
large enough $L$). At a critical point or critical phase, the wave
function is multifractal: it is delocalized in real and momentum-space
and both the ${\rm IPR}$ and ${\rm IPR}_{k}$ scale down with $L$
\citep{Aulbach_2004}.

\paragraph{Exact duality.---}

\begin{figure}[h]
\centering{}\includegraphics[width=1\columnwidth]{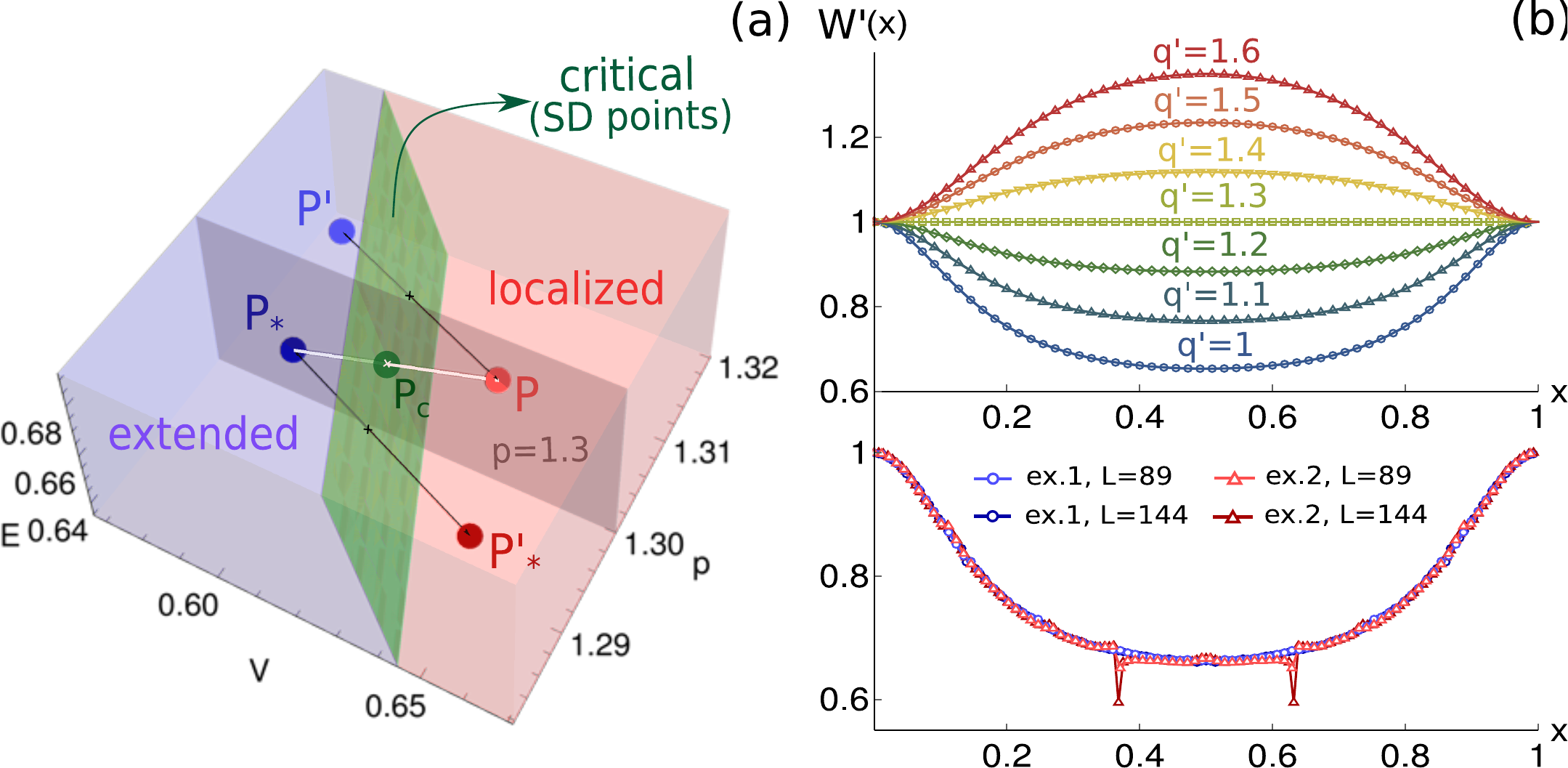}\caption{(a) Example of globally dual points obeying the global duality in
Eq.$\,$\eqref{eq:duality_transformation} (sets of points $P\leftrightarrow P'$
and $P_{*}\leftrightarrow P'_{*}$ connected by black lines), locally
dual points obeying local hidden dualities ($P\leftrightarrow P_{*}$
connected by white line, belonging to plane $p=1.3$) and a SD critical
point $P_{c}$. For this figure, we have set $q=q'=1.$ (b) Top: Duality
function $W'(x)\propto W(2\pi x)$ for a point $P$ defined by $p=1.3,q=1,V\approx0.73,E\approx0.34$,
and different dual points parameterized by different values of $q'$
(the remaining dual parameters, $V',p'$ and $E'$ were obtained by
solving Eq.$\,$\eqref{eq:dual_parameters} for the different choices
of $q'$). The data points correspond to the $L=55$ samples of the
duality function $W'(x)$ obtained for a CA with $\tau_{c}=34/55$
(see \citep{SM} for details). The full lines are plots of the exact
analytical duality function in Eq.$\,$\eqref{eq:duality_transformation}.
The latter was normalized so that $W(0)=W'(0)$. Bottom: Examples
of samples of $W'(x)$ for different locally dual points within the
plane $p=1.3$, for $\tau_{c}=55/89$ and $\tau_{c}=89/144$ (the
energies for the different CA were chosen to be the closest possible
to each other). \label{fig:duality_transformations}}
\end{figure}

The Schrödinger equation for the model in Eq.$\,$\eqref{eq:model-k}
with phase twists $k$ can be written as

\begin{equation}
h_{n}u_{n}-\sum_{m=-\infty}^{\infty}e^{i(\alpha-k)(n-m)}e^{-p|n-m|}u_{m}=0\,,\label{eq:schrodinger_main}
\end{equation}
where $h_{n}=\eta-V\chi(q,2\pi\tau n+\phi)$, $\eta=E+t+V$ and $\chi(\lambda,x)=\sum_{l}e^{-\lambda|l|}e^{ilx}=\sinh(\lambda)[\cosh(\lambda)-\cos(x)]^{-1}$.
At dual points $P(t,V,p,q,\alpha,E;\phi,k)$ and $P'(t',V',p',q',\alpha',E';\phi',k')$,
this equation can be mapped into a dual equation under the duality
transformation (see \citep{SM} for proof):

\begin{equation}
\tilde{u}_{n}=\sum_{m}e^{i2\pi\tau nm}W(2\pi\tau m)u_{m},\label{eq:duality_transformation}
\end{equation}
where $W(x)=\chi(q',x+\phi')\chi^{-1}(p,x+k-\alpha)$. The dual points
$P$ and $P'$ satisfy

\begin{equation}
\begin{cases}
\phi'=k-\alpha+\pi\frac{(s-1)}{2},\textrm{ }-k'+\alpha'=\phi+\pi\frac{(s-1)}{2}\\
\frac{D(V',\eta',p',q')}{B(V',\eta',p')}=s\frac{D(V,\eta,p,q)}{A(V,\eta,q)};\textrm{ }\frac{A(V',\eta',q')}{B(V',\eta',p')}=\frac{B(V,\eta,p)}{A(V,\eta,q)}\\
\frac{\eta'}{B(V',\eta',p')}=s\frac{\eta}{A(V,\eta,q)}
\end{cases}\label{eq:dual_parameters}
\end{equation}
where $s=\pm1$ and

\begin{equation}
\begin{aligned}A(V,\eta,q)= & -\eta\cosh q+V\sinh q\\
B(V,\eta,p)= & -\eta\cosh p+t\sinh p\\
D(V,\eta,p,q)= & \eta\cosh p\cosh q-t\cosh q\sinh p\\
 & -V\cosh p\sinh q\,.
\end{aligned}
\label{eq:coefs}
\end{equation}
For fixed $p=q$, Eq.$\,$\eqref{eq:duality_transformation} defines
the usual Aubry-André duality. The self-duality condition is imposed
by choosing $P=P'$. In this case, Eq.$\,$\eqref{eq:dual_parameters}
is solved simply through the condition $A(V,\eta,q)=\pm B(V,\eta,p)$,
that yields the following equation for the SD points:

\begin{equation}
E=\frac{V\sinh q\mp t\sinh p}{\cosh q\mp\cosh p}-t-V.\label{eq:SD_points}
\end{equation}
Examples of dual points are shown in Fig.$\,$\ref{fig:duality_transformations}(a).
Points $P$ and $P'$ are globally dual, being described by the duality
transformation in Eq.$\,$\eqref{eq:duality_transformation}, as well
as points $P^{*}$ and $P^{'*}$. However, local dualities can also
arise close to the SD points even along directions in the parameter
space where the the global duality breaks down \citep{HdualitiesScipost}.
Examples are the points $P^{*}$ and $P$ in Fig.$\,$\ref{fig:duality_transformations}(a).
These locally dual points are defined by invariant local energy dispersions
under the interchange $k\leftrightarrow\phi$, but only for large
enough $L$ \citep{HdualitiesScipost}.

The global duality transformation defined in Eq.$\,$\eqref{eq:duality_transformation}
was confirmed to match the definition given in Ref.~\citep{HdualitiesScipost}
in terms of CA. Given dual points in the phase diagram, the definition
introduced in Ref.~\citep{HdualitiesScipost} allows, for a given
CA, to calculate $L$ samples of the associated duality function $W'(x)\propto W(2\pi x)$
at points $x_{n}=\mod(\tau_{c}n/,1),n=0,\cdots,L-1$ (see \citep{SM},
Sec-$\,$S2 for details). Fig.$\,$\ref{fig:duality_transformations}(b-top)
shows perfect agreement between the exact global duality function
$W(x)$ in Eq.$\,$\eqref{eq:duality_transformation} and the samples
of duality function $W'(x)$ computed through a CA with $L=55$ sites.
The results were obtained by choosing a fixed point $P$ and different
dual points $P'$ defined by varying $q'$. This illustrates that
even though $P$ is fixed, the duality transformation depends on its
dual point $P'$ (in particular on $q'$), in accordance with the
definition in Eq.$\,$\eqref{eq:duality_transformation}, a feature
that is absent in previously found exact duality transformations \citep{PhysRevLett.104.070601,PhysRevLett.114.146601}.
Finally, Fig.$\,$\ref{fig:duality_transformations}(b-bottom) shows
examples of duality functions $W'(x)$ obtained at locally dual points,
being non-smooth for specific values of $x$, as previously found
for other models \citep{HdualitiesScipost} \footnote{The local duality function may be very sensitive to the choice of
the correct dual points. Since their computation was done numerically,
there is an associated error. Therefore we slightly varied the computed
dual point and checked that some features may arise only due to a
slightly incorrect choice of this point (see \citep{SM}).}.

\begin{figure}[h]
\centering{}\includegraphics[width=1\columnwidth]{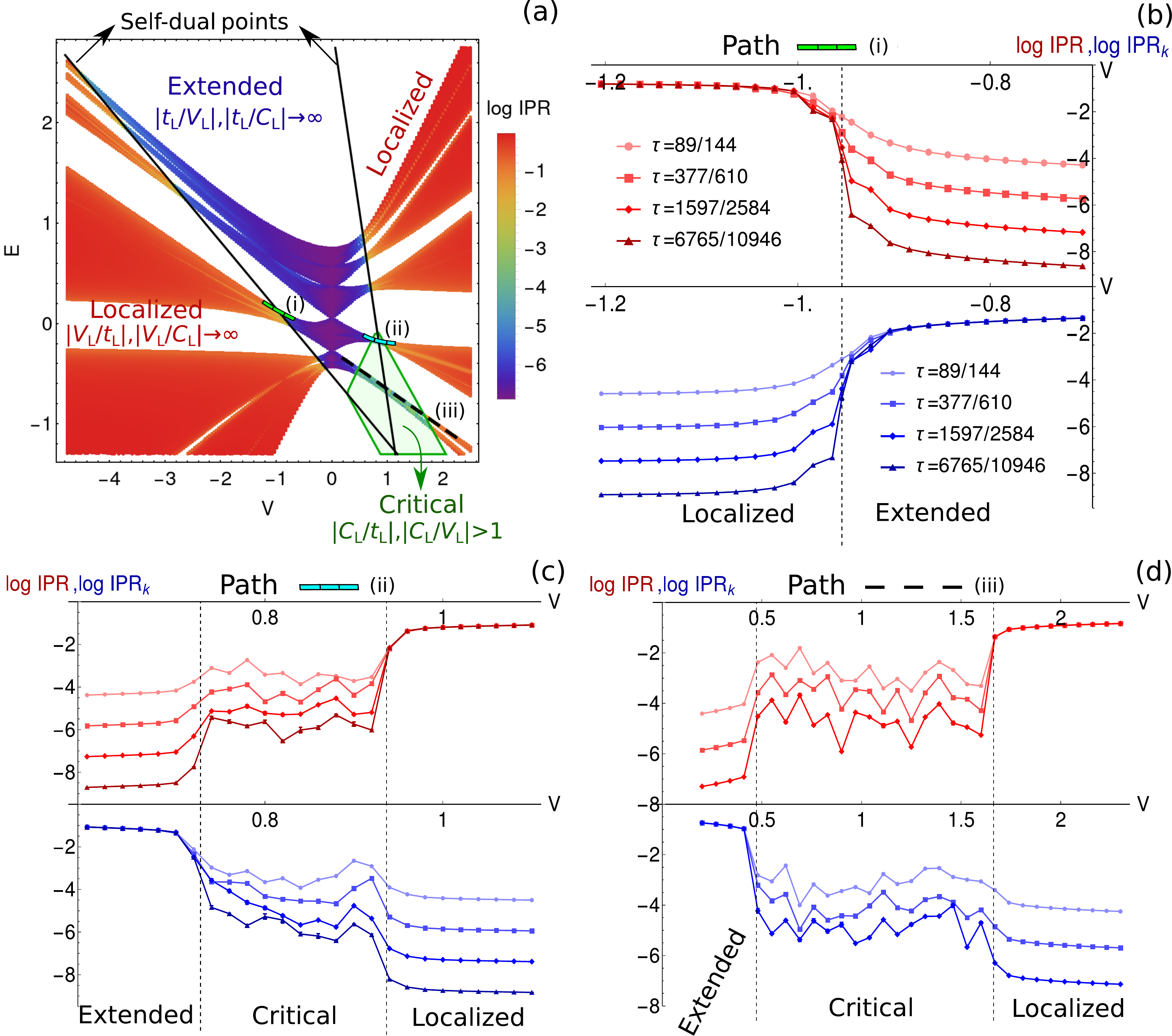}\caption{(a) IPR results obtained for $p=1.3,q=1$ and $L=F_{16}=987$, superimposed
with the analytical curves for SD points (black) and phase boundaries
of the critical phase (green). In each phase, we also show the asymptotic
results of the renormalized couplings as $L\rightarrow\infty$. (b-d)
Finite-size scalings of the ${\rm IPR}$ (red) and ${\rm IPR_{k}}$
(blue) for points $(V,E)$ across the paths shown in the dashed curves
in (a). The results were averaged over $70,25,16$ and $6$ random
shifts $\phi$ and twists $k$, respectively for increasing $L\in[144,10946]$.
The dashed vertical lines correspond to the analytical results for
the phase boundaries. \label{fig:finite_size_scalings}}
\end{figure}

\paragraph*{Phase diagram.---}

We now analytically obtain the complete phase diagram. The transitions
between extended and localized phases obtained through the ${\rm IPR/IPR_{k}}$
calculations perfectly match the SD points described by Eq.$\,$\eqref{eq:SD_points}.
Examples are shown in Fig.~\ref{fig:dualities}(a) for $p=q$, when
Eq.~\eqref{eq:SD_points} reduces to the Aubry-André energy independent
SD line $V=t$, and Fig.$\,$\ref{fig:finite_size_scalings}(a,b)
for $p\neq q$. However, the SD points can also occur within the critical
phase, in which case they are not associated with any transition.
This implies that the phase boundaries of the critical phase are not
described by SD points.

To obtain the full phase diagram analytically we make use of the renormalization-group
approach developed in Ref.~\citep{GoncalvesRG2022}. In fact, the
model studied here is a fixed-point model according to the classification
in \citep{GoncalvesRG2022}. Its characteristic polynomial $\mathcal{P}_{L}\left(\varphi,\kappa\right)\equiv\det[H_{L}(\varphi,\kappa)-E]$,
with $H_{L}(\varphi,\kappa)$ the Hamiltonian for a CA with $L$ sites,
is (see \citep{SM}):

\noindent {\small{}
\begin{equation}
\begin{aligned}\mathcal{P}_{L}(\varphi,\kappa)=V_{L}\cos(\varphi)+t_{L}\cos(\kappa)+C_{L}\cos(\varphi)\cos(\kappa)+D_{L}\end{aligned}
\end{equation}
}where $\varphi=L\phi$, $\kappa=Lk$ and $V_{L},t_{L},C_{L}$ and
$D_{L}$ are renormalized couplings. For the simplest CA (one site
per unit cell), we have, using the definitions in Eq.$\,$\eqref{eq:coefs},
that $t_{1}=A(V,\eta,q)$, $V_{1}=B(V,\eta,p)$, $C_{1}=\eta$ and
$D_{1}=D(V,\eta,p,q)$. The ratios between the renormalized couplings
$V_{L},t_{L}$ and $C_{L}$ can be computed exactly. If $|t_{1}/C_{1}|>1$
or $|V_{1}/C_{1}|>1$, we have, respectively

{\small{}
\begin{equation}
\begin{aligned}\Big|\frac{t_{L}}{C_{L}}\Big|= & \frac{\Big|g_{L}^{+}\Big(\frac{t_{1}}{C_{1}}\Big)+g_{L}^{-}\Big(\frac{t_{1}}{C_{1}}\Big)\Big|}{2}\textrm{ },\textrm{ }\Big|\frac{V_{L}}{C_{L}}\Big|=\frac{\Big|g_{L}^{+}\Big(\frac{V_{1}}{C_{1}}\Big)+g_{L}^{-}\Big(\frac{V_{1}}{C_{1}}\Big)\Big|}{2}\end{aligned}
\end{equation}
}where $g_{L}^{\pm}(x)=\Big(x\pm\sqrt{x^{2}-1}\Big)^{L}$. On the
other hand, if $|C_{1}/t_{1}|>1$ or $|C_{1}/V_{1}|>1$ we have, respectively

\begin{equation}
|t_{L}/C_{L}|=|T_{L}(t_{1}/C_{1})|;\textrm{ }|V_{L}/C_{L}|=|T_{L}(V_{1}/C_{1})|,
\end{equation}
where $T_{L}(x)$ is the $L$-th order Chebyshev polynomial. It is
easy to see that if $|t_{1}/V_{1}|,|t_{1}/C_{1}|>1$ we have that
$|t_{L}/V_{L}|,|t_{L}/C_{L}|\rightarrow\infty$ exponentially in $L$
as $L\rightarrow\infty$, i.e. we are in the extended phase. For $|V_{1}/t_{1}|,|V_{1}/C_{1}|>1$,
we have $|V_{L}/t_{L}|,|V_{L}/C_{L}|\rightarrow\infty$ and the phase
is localized. Finally, $|C_{1}/t_{1}|,|C_{1}/V_{1}|>1$ ensures that
$|C_{L}/t_{L}|,|C_{L}/V_{L}|>1$ for any $L$ (a property of Chebyshev
polynomials), and the system is in a critical phase. Therefore, the
phases and phase boundaries are fully determined through the previous
conditions by knowing the functions in Eq.$\,$\eqref{eq:coefs}.
Summarizing, phases and phase boundaries are analytically given by

\begin{equation}
\begin{aligned}|A/B|,|A/\eta|>1 & \textrm{, Ext.}\\
|B/A|,|B/\eta|>1 & \textrm{, Loc.}\\
|\eta/A|,|\eta/B|>1 & \textrm{, Crit.}
\end{aligned}
\label{eq:phases}
\end{equation}

\vspace{-0.5cm}

\begin{equation}
\begin{aligned}|A|=|B|,\textrm{ }|A|,|B|>|\eta| & \textrm{, Ext.-to-Loc.}\\
|A|=|\eta|,\textrm{ }|A|,|\eta|>|B| & \textrm{, Crit.-to-Ext.}\\
|B|=|\eta|,\textrm{ }|B|,|\eta|>|A| & \textrm{, Crit.-to-Loc.}
\end{aligned}
\label{eq:phase_boundaries}
\end{equation}
where we omitted the parameter dependence for clarity. From the ratios
of renormalized couplings we are also able to calculate the correlation
lengths in the extended and localized phases in terms of $A,B$ and
$\eta$ (see \citep{SM}). Note that the $L\rightarrow\infty$ limit
defines the phase diagram for any $\tau$ because the renormalized
couplings only depend on $L$.

To confirm our analytical results, we show in Figs.$\,$\ref{fig:finite_size_scalings}(b-d)
some examples of finite-size scaling results that agree with the analytical
phase boundaries here unveiled. Note that while in the extended-to-localized
transitions both the ${\rm IPR}$ and ${\rm IPR}_{k}$ scale down
only at the critical point {[}Fig.$\,$\ref{fig:finite_size_scalings}(b){]},
such scaling is observed for the entire range of the critical phase
when the latter exists {[}Fig.$\,$\ref{fig:finite_size_scalings}(c-d){]}.
In \citep{SM} we also carried out a multifractal analysis at some
points in the critical phase to show the non-linear behaviour of the
fractal dimension that characterizes multifractal phases \citep{Janssen}.

\paragraph{Discussion.---}

We analytically obtained the phase diagram of the richest family of
1D quasiperiodic solvable models, to our knowledge, hosting (i) critical
multifractal phases in addition to localized and extended ones, and
energy-dependent transitions between all these phases; and (ii) a
rich generalized duality symmetry that includes dualities inside the
critical phase.

From a practical perspective, the family of models we propose can
be experimentally realized with currently available techniques. The
model in \citep{PhysRevLett.114.146601} have exactly the quasiperiodic
potential considered here and was already experimentally realized
using a synthetic lattice of laser-coupled momentum modes \citep{PhysRevLett.126.040603}.
Our model simply requires additional longer-range hoppings (but still
exponentially-decaying), a possibility put forward in \citep{PhysRevA.92.043606}.
It can also be simulated in conventional optical lattices, where the
exponential hopping decay rate can be directly estimated \citep{PhysRevLett.104.070601}.
A single incommensurate potential (our large $q$ limit) was already
realized in optical lattices by applying a second laser beam with
a wave vector $\tau$ that is incommensurate with that of the primary
lattice. Additional quasiperiodic harmonics can be introduced by adding
new laser beams with wave vectors that are multiples of $\tau$, as
proposed in \citep{PhysRevLett.114.146601}. The engineering of optical
lattices with kicked kinetic energy or quasiperiodic potential is
also a possible way to implement our model, as proposed in \citep{anomScipost}.
An advantage of our model is that the critical multifractal phase
can be realized without the need of unbounded potentials. We also
note that the existence of exact dualities has direct experimental
relevance: critical-extended transitions are dual of critical-localized
transitions, implying that the detailed experimental characterization
of one transition can give us information on both. The impact of interactions
on the phase diagram of this model is an interesting question for
future research.

\label{sec:Discussion}

$\textrm{ }$
\begin{acknowledgments}
The authors MG and PR acknowledge partial support from Fundação para
a Ciência e Tecnologia (FCT-Portugal) through Grant No. UID/CTM/04540/2019.
BA and EVC acknowledge partial support from FCT-Portugal through Grant
No. UIDB/04650/2020. MG acknowledges further support from FCT-Portugal
through the Grant SFRH/BD/145152/2019. BA acknowledges further support
from FCT-Portugal through Grant No. CEECIND/02936/2017. We finally
acknowledge the Tianhe-2JK cluster at the Beijing Computational Science
Research Center (CSRC) and the OBLIVION supercomputer (based at the
High Performance Computing Center - University of Évora) funded by
the ENGAGE SKA Research Infrastructure (reference POCI-01-0145-FEDER-022217
- COMPETE 2020 and the Foundation for Science and Technology, Portugal)
and by the BigData@UE project (reference ALT20-03-0246-FEDER-000033
- FEDER) and the Alentejo 2020 Regional Operational Program. Computer
assistance was provided by CSRC and the OBLIVION support team.
\end{acknowledgments}

\bibliographystyle{apsrev4-1}
\bibliography{1D_Hidden_SD_Paper}

\begin{thebibliography}{76}%
\makeatletter
\providecommand \@ifxundefined [1]{%
 \@ifx{#1\undefined}
}%
\providecommand \@ifnum [1]{%
 \ifnum #1\expandafter \@firstoftwo
 \else \expandafter \@secondoftwo
 \fi
}%
\providecommand \@ifx [1]{%
 \ifx #1\expandafter \@firstoftwo
 \else \expandafter \@secondoftwo
 \fi
}%
\providecommand \natexlab [1]{#1}%
\providecommand \enquote  [1]{``#1''}%
\providecommand \bibnamefont  [1]{#1}%
\providecommand \bibfnamefont [1]{#1}%
\providecommand \citenamefont [1]{#1}%
\providecommand \href@noop [0]{\@secondoftwo}%
\providecommand \href [0]{\begingroup \@sanitize@url \@href}%
\providecommand \@href[1]{\@@startlink{#1}\@@href}%
\providecommand \@@href[1]{\endgroup#1\@@endlink}%
\providecommand \@sanitize@url [0]{\catcode `\\12\catcode `\$12\catcode
  `\&12\catcode `\#12\catcode `\^12\catcode `\_12\catcode `\%12\relax}%
\providecommand \@@startlink[1]{}%
\providecommand \@@endlink[0]{}%
\providecommand \url  [0]{\begingroup\@sanitize@url \@url }%
\providecommand \@url [1]{\endgroup\@href {#1}{\urlprefix }}%
\providecommand \urlprefix  [0]{URL }%
\providecommand \Eprint [0]{\href }%
\providecommand \doibase [0]{http://dx.doi.org/}%
\providecommand \selectlanguage [0]{\@gobble}%
\providecommand \bibinfo  [0]{\@secondoftwo}%
\providecommand \bibfield  [0]{\@secondoftwo}%
\providecommand \translation [1]{[#1]}%
\providecommand \BibitemOpen [0]{}%
\providecommand \bibitemStop [0]{}%
\providecommand \bibitemNoStop [0]{.\EOS\space}%
\providecommand \EOS [0]{\spacefactor3000\relax}%
\providecommand \BibitemShut  [1]{\csname bibitem#1\endcsname}%
\let\auto@bib@innerbib\@empty
\bibitem [{\citenamefont {Aubry}\ and\ \citenamefont
  {Andr{\'{e}}}(1980)}]{AubryAndre}%
  \BibitemOpen
  \bibfield  {author} {\bibinfo {author} {\bibfnamefont {S.}~\bibnamefont
  {Aubry}}\ and\ \bibinfo {author} {\bibfnamefont {G.}~\bibnamefont
  {Andr{\'{e}}}},\ }\href@noop {} {\bibfield  {journal} {\bibinfo  {journal}
  {Proceedings, VIII International Colloquium on Group-Theoretical Methods in
  Physics}\ }\textbf {\bibinfo {volume} {3}} (\bibinfo {year}
  {1980})}\BibitemShut {NoStop}%
\bibitem [{\citenamefont {Roati}\ \emph {et~al.}(2008)\citenamefont {Roati},
  \citenamefont {D'Errico}, \citenamefont {Fallani}, \citenamefont {Fattori},
  \citenamefont {Fort}, \citenamefont {Zaccanti}, \citenamefont {Modugno},
  \citenamefont {Modugno},\ and\ \citenamefont {Inguscio}}]{Roati2008}%
  \BibitemOpen
  \bibfield  {author} {\bibinfo {author} {\bibfnamefont {G.}~\bibnamefont
  {Roati}}, \bibinfo {author} {\bibfnamefont {C.}~\bibnamefont {D'Errico}},
  \bibinfo {author} {\bibfnamefont {L.}~\bibnamefont {Fallani}}, \bibinfo
  {author} {\bibfnamefont {M.}~\bibnamefont {Fattori}}, \bibinfo {author}
  {\bibfnamefont {C.}~\bibnamefont {Fort}}, \bibinfo {author} {\bibfnamefont
  {M.}~\bibnamefont {Zaccanti}}, \bibinfo {author} {\bibfnamefont
  {G.}~\bibnamefont {Modugno}}, \bibinfo {author} {\bibfnamefont
  {M.}~\bibnamefont {Modugno}}, \ and\ \bibinfo {author} {\bibfnamefont
  {M.}~\bibnamefont {Inguscio}},\ }\href {\doibase 10.1038/nature07071}
  {\bibfield  {journal} {\bibinfo  {journal} {Nature}\ }\textbf {\bibinfo
  {volume} {453}},\ \bibinfo {pages} {895} (\bibinfo {year} {2008})},\ \Eprint
  {http://arxiv.org/abs/0804.2609} {arXiv:0804.2609} \BibitemShut {NoStop}%
\bibitem [{\citenamefont {Lahini}\ \emph {et~al.}(2009)\citenamefont {Lahini},
  \citenamefont {Pugatch}, \citenamefont {Pozzi}, \citenamefont {Sorel},
  \citenamefont {Morandotti}, \citenamefont {Davidson},\ and\ \citenamefont
  {Silberberg}}]{Lahini2009}%
  \BibitemOpen
  \bibfield  {author} {\bibinfo {author} {\bibfnamefont {Y.}~\bibnamefont
  {Lahini}}, \bibinfo {author} {\bibfnamefont {R.}~\bibnamefont {Pugatch}},
  \bibinfo {author} {\bibfnamefont {F.}~\bibnamefont {Pozzi}}, \bibinfo
  {author} {\bibfnamefont {M.}~\bibnamefont {Sorel}}, \bibinfo {author}
  {\bibfnamefont {R.}~\bibnamefont {Morandotti}}, \bibinfo {author}
  {\bibfnamefont {N.}~\bibnamefont {Davidson}}, \ and\ \bibinfo {author}
  {\bibfnamefont {Y.}~\bibnamefont {Silberberg}},\ }\href {\doibase
  10.1103/PhysRevLett.103.013901} {\bibfield  {journal} {\bibinfo  {journal}
  {Phys. Rev. Lett.}\ }\textbf {\bibinfo {volume} {103}},\ \bibinfo {pages}
  {013901} (\bibinfo {year} {2009})}\BibitemShut {NoStop}%
\bibitem [{\citenamefont {Schreiber}\ \emph {et~al.}(2015)\citenamefont
  {Schreiber}, \citenamefont {Hodgman}, \citenamefont {Bordia}, \citenamefont
  {L{\"{u}}schen}, \citenamefont {Fischer}, \citenamefont {Vosk}, \citenamefont
  {Altman}, \citenamefont {Schneider},\ and\ \citenamefont
  {Bloch}}]{Schreiber842}%
  \BibitemOpen
  \bibfield  {author} {\bibinfo {author} {\bibfnamefont {M.}~\bibnamefont
  {Schreiber}}, \bibinfo {author} {\bibfnamefont {S.~S.}\ \bibnamefont
  {Hodgman}}, \bibinfo {author} {\bibfnamefont {P.}~\bibnamefont {Bordia}},
  \bibinfo {author} {\bibfnamefont {H.~P.}\ \bibnamefont {L{\"{u}}schen}},
  \bibinfo {author} {\bibfnamefont {M.~H.}\ \bibnamefont {Fischer}}, \bibinfo
  {author} {\bibfnamefont {R.}~\bibnamefont {Vosk}}, \bibinfo {author}
  {\bibfnamefont {E.}~\bibnamefont {Altman}}, \bibinfo {author} {\bibfnamefont
  {U.}~\bibnamefont {Schneider}}, \ and\ \bibinfo {author} {\bibfnamefont
  {I.}~\bibnamefont {Bloch}},\ }\href {\doibase 10.1126/science.aaa7432}
  {\bibfield  {journal} {\bibinfo  {journal} {Science}\ }\textbf {\bibinfo
  {volume} {349}},\ \bibinfo {pages} {842} (\bibinfo {year} {2015})},\ \Eprint
  {http://arxiv.org/abs/1501.05661} {arXiv:1501.05661} \BibitemShut {NoStop}%
\bibitem [{\citenamefont {L\"uschen}\ \emph {et~al.}(2018)\citenamefont
  {L\"uschen}, \citenamefont {Scherg}, \citenamefont {Kohlert}, \citenamefont
  {Schreiber}, \citenamefont {Bordia}, \citenamefont {Li}, \citenamefont
  {Das~Sarma},\ and\ \citenamefont {Bloch}}]{Luschen2018}%
  \BibitemOpen
  \bibfield  {author} {\bibinfo {author} {\bibfnamefont {H.~P.}\ \bibnamefont
  {L\"uschen}}, \bibinfo {author} {\bibfnamefont {S.}~\bibnamefont {Scherg}},
  \bibinfo {author} {\bibfnamefont {T.}~\bibnamefont {Kohlert}}, \bibinfo
  {author} {\bibfnamefont {M.}~\bibnamefont {Schreiber}}, \bibinfo {author}
  {\bibfnamefont {P.}~\bibnamefont {Bordia}}, \bibinfo {author} {\bibfnamefont
  {X.}~\bibnamefont {Li}}, \bibinfo {author} {\bibfnamefont {S.}~\bibnamefont
  {Das~Sarma}}, \ and\ \bibinfo {author} {\bibfnamefont {I.}~\bibnamefont
  {Bloch}},\ }\href {\doibase 10.1103/PhysRevLett.120.160404} {\bibfield
  {journal} {\bibinfo  {journal} {Phys. Rev. Lett.}\ }\textbf {\bibinfo
  {volume} {120}},\ \bibinfo {pages} {160404} (\bibinfo {year}
  {2018})}\BibitemShut {NoStop}%
\bibitem [{\citenamefont {Borgnia}\ \emph {et~al.}(2022)\citenamefont
  {Borgnia}, \citenamefont {Vishwanath},\ and\ \citenamefont
  {Slager}}]{slager1}%
  \BibitemOpen
  \bibfield  {author} {\bibinfo {author} {\bibfnamefont {D.~S.}\ \bibnamefont
  {Borgnia}}, \bibinfo {author} {\bibfnamefont {A.}~\bibnamefont {Vishwanath}},
  \ and\ \bibinfo {author} {\bibfnamefont {R.-J.}\ \bibnamefont {Slager}},\
  }\href {\doibase 10.1103/PhysRevB.106.054204} {\bibfield  {journal} {\bibinfo
   {journal} {Phys. Rev. B}\ }\textbf {\bibinfo {volume} {106}},\ \bibinfo
  {pages} {054204} (\bibinfo {year} {2022})}\BibitemShut {NoStop}%
\bibitem [{\citenamefont {Huang}\ \emph {et~al.}(2016)\citenamefont {Huang},
  \citenamefont {Ye}, \citenamefont {Chen}, \citenamefont {Kartashov},
  \citenamefont {Konotop},\ and\ \citenamefont {Torner}}]{Huang2016a}%
  \BibitemOpen
  \bibfield  {author} {\bibinfo {author} {\bibfnamefont {C.}~\bibnamefont
  {Huang}}, \bibinfo {author} {\bibfnamefont {F.}~\bibnamefont {Ye}}, \bibinfo
  {author} {\bibfnamefont {X.}~\bibnamefont {Chen}}, \bibinfo {author}
  {\bibfnamefont {Y.~V.}\ \bibnamefont {Kartashov}}, \bibinfo {author}
  {\bibfnamefont {V.~V.}\ \bibnamefont {Konotop}}, \ and\ \bibinfo {author}
  {\bibfnamefont {L.}~\bibnamefont {Torner}},\ }\href {\doibase
  10.1038/srep32546} {\bibfield  {journal} {\bibinfo  {journal} {Scientific
  Reports}\ }\textbf {\bibinfo {volume} {6}},\ \bibinfo {pages} {32546}
  (\bibinfo {year} {2016})}\BibitemShut {NoStop}%
\bibitem [{\citenamefont {Pixley}\ \emph {et~al.}(2018)\citenamefont {Pixley},
  \citenamefont {Wilson}, \citenamefont {Huse},\ and\ \citenamefont
  {Gopalakrishnan}}]{PhysRevLett.120.207604}%
  \BibitemOpen
  \bibfield  {author} {\bibinfo {author} {\bibfnamefont {J.~H.}\ \bibnamefont
  {Pixley}}, \bibinfo {author} {\bibfnamefont {J.~H.}\ \bibnamefont {Wilson}},
  \bibinfo {author} {\bibfnamefont {D.~A.}\ \bibnamefont {Huse}}, \ and\
  \bibinfo {author} {\bibfnamefont {S.}~\bibnamefont {Gopalakrishnan}},\ }\href
  {\doibase 10.1103/PhysRevLett.120.207604} {\bibfield  {journal} {\bibinfo
  {journal} {Phys. Rev. Lett.}\ }\textbf {\bibinfo {volume} {120}},\ \bibinfo
  {pages} {207604} (\bibinfo {year} {2018})}\BibitemShut {NoStop}%
\bibitem [{\citenamefont {Park}\ \emph {et~al.}(2019)\citenamefont {Park},
  \citenamefont {Kim},\ and\ \citenamefont {Lee}}]{Park2018}%
  \BibitemOpen
  \bibfield  {author} {\bibinfo {author} {\bibfnamefont {M.~J.}\ \bibnamefont
  {Park}}, \bibinfo {author} {\bibfnamefont {H.~S.}\ \bibnamefont {Kim}}, \
  and\ \bibinfo {author} {\bibfnamefont {S.}~\bibnamefont {Lee}},\ }\href
  {\doibase 10.1103/PhysRevB.99.245401} {\bibfield  {journal} {\bibinfo
  {journal} {Phys. Rev. B}\ }\textbf {\bibinfo {volume} {99}},\ \bibinfo
  {pages} {245401} (\bibinfo {year} {2019})},\ \Eprint
  {http://arxiv.org/abs/1812.09170} {arXiv:1812.09170} \BibitemShut {NoStop}%
\bibitem [{\citenamefont {Huang}\ and\ \citenamefont
  {Liu}(2019)}]{PhysRevB.100.144202}%
  \BibitemOpen
  \bibfield  {author} {\bibinfo {author} {\bibfnamefont {B.}~\bibnamefont
  {Huang}}\ and\ \bibinfo {author} {\bibfnamefont {W.~V.}\ \bibnamefont
  {Liu}},\ }\href {\doibase 10.1103/PhysRevB.100.144202} {\bibfield  {journal}
  {\bibinfo  {journal} {Phys. Rev. B}\ }\textbf {\bibinfo {volume} {100}},\
  \bibinfo {pages} {144202} (\bibinfo {year} {2019})}\BibitemShut {NoStop}%
\bibitem [{\citenamefont {Fu}\ \emph {et~al.}(2020)\citenamefont {Fu},
  \citenamefont {K{\"{o}}nig}, \citenamefont {Wilson}, \citenamefont {Chou},\
  and\ \citenamefont {Pixley}}]{Fu2020}%
  \BibitemOpen
  \bibfield  {author} {\bibinfo {author} {\bibfnamefont {Y.}~\bibnamefont
  {Fu}}, \bibinfo {author} {\bibfnamefont {E.~J.}\ \bibnamefont {K{\"{o}}nig}},
  \bibinfo {author} {\bibfnamefont {J.~H.}\ \bibnamefont {Wilson}}, \bibinfo
  {author} {\bibfnamefont {Y.-Z.}\ \bibnamefont {Chou}}, \ and\ \bibinfo
  {author} {\bibfnamefont {J.~H.}\ \bibnamefont {Pixley}},\ }\href {\doibase
  10.1038/s41535-020-00271-9} {\bibfield  {journal} {\bibinfo  {journal} {npj
  Quantum Materials}\ }\textbf {\bibinfo {volume} {5}},\ \bibinfo {pages} {71}
  (\bibinfo {year} {2020})}\BibitemShut {NoStop}%
\bibitem [{\citenamefont {Wang}\ \emph
  {et~al.}(2020{\natexlab{a}})\citenamefont {Wang}, \citenamefont {Zheng},
  \citenamefont {Chen}, \citenamefont {Huang}, \citenamefont {Kartashov},
  \citenamefont {Torner}, \citenamefont {Konotop},\ and\ \citenamefont
  {Ye}}]{Wang2020}%
  \BibitemOpen
  \bibfield  {author} {\bibinfo {author} {\bibfnamefont {P.}~\bibnamefont
  {Wang}}, \bibinfo {author} {\bibfnamefont {Y.}~\bibnamefont {Zheng}},
  \bibinfo {author} {\bibfnamefont {X.}~\bibnamefont {Chen}}, \bibinfo {author}
  {\bibfnamefont {C.}~\bibnamefont {Huang}}, \bibinfo {author} {\bibfnamefont
  {Y.~V.}\ \bibnamefont {Kartashov}}, \bibinfo {author} {\bibfnamefont
  {L.}~\bibnamefont {Torner}}, \bibinfo {author} {\bibfnamefont {V.~V.}\
  \bibnamefont {Konotop}}, \ and\ \bibinfo {author} {\bibfnamefont
  {F.}~\bibnamefont {Ye}},\ }\href {\doibase 10.1038/s41586-019-1851-6}
  {\bibfield  {journal} {\bibinfo  {journal} {Nature}\ }\textbf {\bibinfo
  {volume} {577}},\ \bibinfo {pages} {42} (\bibinfo {year}
  {2020}{\natexlab{a}})}\BibitemShut {NoStop}%
\bibitem [{\citenamefont {Gon{\c{c}}alves}\ \emph {et~al.}(2020)\citenamefont
  {Gon{\c{c}}alves}, \citenamefont {Olyaei}, \citenamefont {Amorim},
  \citenamefont {Mondaini}, \citenamefont {Ribeiro},\ and\ \citenamefont
  {Castro}}]{goncalves2020}%
  \BibitemOpen
  \bibfield  {author} {\bibinfo {author} {\bibfnamefont {M.}~\bibnamefont
  {Gon{\c{c}}alves}}, \bibinfo {author} {\bibfnamefont {H.~Z.}\ \bibnamefont
  {Olyaei}}, \bibinfo {author} {\bibfnamefont {B.}~\bibnamefont {Amorim}},
  \bibinfo {author} {\bibfnamefont {R.}~\bibnamefont {Mondaini}}, \bibinfo
  {author} {\bibfnamefont {P.}~\bibnamefont {Ribeiro}}, \ and\ \bibinfo
  {author} {\bibfnamefont {E.~V.}\ \bibnamefont {Castro}},\ }\href@noop {}
  {\enquote {\bibinfo {title} {{Incommensurability-induced sub-ballistic
  narrow-band-states in twisted bilayer graphene}},}\ } (\bibinfo {year}
  {2020}),\ \Eprint {http://arxiv.org/abs/2008.07542} {arXiv:2008.07542
  [cond-mat.mes-hall]} \BibitemShut {NoStop}%
\bibitem [{\citenamefont {Bordia}\ \emph
  {et~al.}(2017{\natexlab{a}})\citenamefont {Bordia}, \citenamefont
  {L\"uschen}, \citenamefont {Scherg}, \citenamefont {Gopalakrishnan},
  \citenamefont {Knap}, \citenamefont {Schneider},\ and\ \citenamefont
  {Bloch}}]{PhysRevX.7.041047}%
  \BibitemOpen
  \bibfield  {author} {\bibinfo {author} {\bibfnamefont {P.}~\bibnamefont
  {Bordia}}, \bibinfo {author} {\bibfnamefont {H.}~\bibnamefont {L\"uschen}},
  \bibinfo {author} {\bibfnamefont {S.}~\bibnamefont {Scherg}}, \bibinfo
  {author} {\bibfnamefont {S.}~\bibnamefont {Gopalakrishnan}}, \bibinfo
  {author} {\bibfnamefont {M.}~\bibnamefont {Knap}}, \bibinfo {author}
  {\bibfnamefont {U.}~\bibnamefont {Schneider}}, \ and\ \bibinfo {author}
  {\bibfnamefont {I.}~\bibnamefont {Bloch}},\ }\href {\doibase
  10.1103/PhysRevX.7.041047} {\bibfield  {journal} {\bibinfo  {journal} {Phys.
  Rev. X}\ }\textbf {\bibinfo {volume} {7}},\ \bibinfo {pages} {041047}
  (\bibinfo {year} {2017}{\natexlab{a}})}\BibitemShut {NoStop}%
\bibitem [{\citenamefont {Kraus}\ \emph {et~al.}(2012)\citenamefont {Kraus},
  \citenamefont {Lahini}, \citenamefont {Ringel}, \citenamefont {Verbin},\ and\
  \citenamefont {Zilberberg}}]{Kraus2012}%
  \BibitemOpen
  \bibfield  {author} {\bibinfo {author} {\bibfnamefont {Y.~E.}\ \bibnamefont
  {Kraus}}, \bibinfo {author} {\bibfnamefont {Y.}~\bibnamefont {Lahini}},
  \bibinfo {author} {\bibfnamefont {Z.}~\bibnamefont {Ringel}}, \bibinfo
  {author} {\bibfnamefont {M.}~\bibnamefont {Verbin}}, \ and\ \bibinfo {author}
  {\bibfnamefont {O.}~\bibnamefont {Zilberberg}},\ }\href {\doibase
  10.1103/PhysRevLett.109.106402} {\bibfield  {journal} {\bibinfo  {journal}
  {Phys. Rev. Lett.}\ }\textbf {\bibinfo {volume} {109}},\ \bibinfo {pages}
  {106402} (\bibinfo {year} {2012})}\BibitemShut {NoStop}%
\bibitem [{\citenamefont {Kraus}\ and\ \citenamefont
  {Zilberberg}(2012)}]{PhysRevLett.109.116404}%
  \BibitemOpen
  \bibfield  {author} {\bibinfo {author} {\bibfnamefont {Y.~E.}\ \bibnamefont
  {Kraus}}\ and\ \bibinfo {author} {\bibfnamefont {O.}~\bibnamefont
  {Zilberberg}},\ }\href {\doibase 10.1103/PhysRevLett.109.116404} {\bibfield
  {journal} {\bibinfo  {journal} {Phys. Rev. Lett.}\ }\textbf {\bibinfo
  {volume} {109}},\ \bibinfo {pages} {116404} (\bibinfo {year}
  {2012})}\BibitemShut {NoStop}%
\bibitem [{\citenamefont {Verbin}\ \emph {et~al.}(2013)\citenamefont {Verbin},
  \citenamefont {Zilberberg}, \citenamefont {Kraus}, \citenamefont {Lahini},\
  and\ \citenamefont {Silberberg}}]{Verbin2013}%
  \BibitemOpen
  \bibfield  {author} {\bibinfo {author} {\bibfnamefont {M.}~\bibnamefont
  {Verbin}}, \bibinfo {author} {\bibfnamefont {O.}~\bibnamefont {Zilberberg}},
  \bibinfo {author} {\bibfnamefont {Y.~E.}\ \bibnamefont {Kraus}}, \bibinfo
  {author} {\bibfnamefont {Y.}~\bibnamefont {Lahini}}, \ and\ \bibinfo {author}
  {\bibfnamefont {Y.}~\bibnamefont {Silberberg}},\ }\href {\doibase
  10.1103/PhysRevLett.110.076403} {\bibfield  {journal} {\bibinfo  {journal}
  {Phys. Rev. Lett.}\ }\textbf {\bibinfo {volume} {110}},\ \bibinfo {pages}
  {076403} (\bibinfo {year} {2013})}\BibitemShut {NoStop}%
\bibitem [{\citenamefont {Zilberberg}(2021)}]{Zilberberg:21}%
  \BibitemOpen
  \bibfield  {author} {\bibinfo {author} {\bibfnamefont {O.}~\bibnamefont
  {Zilberberg}},\ }\href {\doibase 10.1364/OME.416552} {\bibfield  {journal}
  {\bibinfo  {journal} {Opt. Mater. Express}\ }\textbf {\bibinfo {volume}
  {11}},\ \bibinfo {pages} {1143} (\bibinfo {year} {2021})}\BibitemShut
  {NoStop}%
\bibitem [{\citenamefont {Borgnia}\ and\ \citenamefont
  {Slager}(2021)}]{slager2}%
  \BibitemOpen
  \bibfield  {author} {\bibinfo {author} {\bibfnamefont {D.~S.}\ \bibnamefont
  {Borgnia}}\ and\ \bibinfo {author} {\bibfnamefont {R.-J.}\ \bibnamefont
  {Slager}},\ }\href {\doibase 10.48550/ARXIV.2111.02789} {\enquote {\bibinfo
  {title} {Localization via quasi-periodic bulk-bulk correspondence},}\ }
  (\bibinfo {year} {2021})\BibitemShut {NoStop}%
\bibitem [{\citenamefont {Boers}\ \emph {et~al.}(2007)\citenamefont {Boers},
  \citenamefont {Goedeke}, \citenamefont {Hinrichs},\ and\ \citenamefont
  {Holthaus}}]{PhysRevA.75.063404}%
  \BibitemOpen
  \bibfield  {author} {\bibinfo {author} {\bibfnamefont {D.~J.}\ \bibnamefont
  {Boers}}, \bibinfo {author} {\bibfnamefont {B.}~\bibnamefont {Goedeke}},
  \bibinfo {author} {\bibfnamefont {D.}~\bibnamefont {Hinrichs}}, \ and\
  \bibinfo {author} {\bibfnamefont {M.}~\bibnamefont {Holthaus}},\ }\href
  {\doibase 10.1103/PhysRevA.75.063404} {\bibfield  {journal} {\bibinfo
  {journal} {Phys. Rev. A}\ }\textbf {\bibinfo {volume} {75}},\ \bibinfo
  {pages} {63404} (\bibinfo {year} {2007})}\BibitemShut {NoStop}%
\bibitem [{\citenamefont {Modugno}(2009)}]{Modugno_2009}%
  \BibitemOpen
  \bibfield  {author} {\bibinfo {author} {\bibfnamefont {M.}~\bibnamefont
  {Modugno}},\ }\href {\doibase 10.1088/1367-2630/11/3/033023} {\bibfield
  {journal} {\bibinfo  {journal} {New Journal of Physics}\ }\textbf {\bibinfo
  {volume} {11}},\ \bibinfo {pages} {33023} (\bibinfo {year}
  {2009})}\BibitemShut {NoStop}%
\bibitem [{\citenamefont {Yao}\ \emph {et~al.}(2019)\citenamefont {Yao},
  \citenamefont {Khoudli}, \citenamefont {Bresque},\ and\ \citenamefont
  {Sanchez-Palencia}}]{PhysRevLett.123.070405}%
  \BibitemOpen
  \bibfield  {author} {\bibinfo {author} {\bibfnamefont {H.}~\bibnamefont
  {Yao}}, \bibinfo {author} {\bibfnamefont {H.}~\bibnamefont {Khoudli}},
  \bibinfo {author} {\bibfnamefont {L.}~\bibnamefont {Bresque}}, \ and\
  \bibinfo {author} {\bibfnamefont {L.}~\bibnamefont {Sanchez-Palencia}},\
  }\href {\doibase 10.1103/PhysRevLett.123.070405} {\bibfield  {journal}
  {\bibinfo  {journal} {Phys. Rev. Lett.}\ }\textbf {\bibinfo {volume} {123}},\
  \bibinfo {pages} {070405} (\bibinfo {year} {2019})}\BibitemShut {NoStop}%
\bibitem [{\citenamefont {Yao}\ \emph {et~al.}(2020)\citenamefont {Yao},
  \citenamefont {Giamarchi},\ and\ \citenamefont
  {Sanchez-Palencia}}]{PhysRevLett.125.060401}%
  \BibitemOpen
  \bibfield  {author} {\bibinfo {author} {\bibfnamefont {H.}~\bibnamefont
  {Yao}}, \bibinfo {author} {\bibfnamefont {T.}~\bibnamefont {Giamarchi}}, \
  and\ \bibinfo {author} {\bibfnamefont {L.}~\bibnamefont {Sanchez-Palencia}},\
  }\href {\doibase 10.1103/PhysRevLett.125.060401} {\bibfield  {journal}
  {\bibinfo  {journal} {Phys. Rev. Lett.}\ }\textbf {\bibinfo {volume} {125}},\
  \bibinfo {pages} {060401} (\bibinfo {year} {2020})}\BibitemShut {NoStop}%
\bibitem [{\citenamefont {Gautier}\ \emph {et~al.}(2021)\citenamefont
  {Gautier}, \citenamefont {Yao},\ and\ \citenamefont
  {Sanchez-Palencia}}]{PhysRevLett.126.110401}%
  \BibitemOpen
  \bibfield  {author} {\bibinfo {author} {\bibfnamefont {R.}~\bibnamefont
  {Gautier}}, \bibinfo {author} {\bibfnamefont {H.}~\bibnamefont {Yao}}, \ and\
  \bibinfo {author} {\bibfnamefont {L.}~\bibnamefont {Sanchez-Palencia}},\
  }\href {\doibase 10.1103/PhysRevLett.126.110401} {\bibfield  {journal}
  {\bibinfo  {journal} {Phys. Rev. Lett.}\ }\textbf {\bibinfo {volume} {126}},\
  \bibinfo {pages} {110401} (\bibinfo {year} {2021})}\BibitemShut {NoStop}%
\bibitem [{\citenamefont {An}\ \emph {et~al.}(2021)\citenamefont {An},
  \citenamefont {Padavi\ifmmode~\acute{c}\else \'{c}\fi{}}, \citenamefont
  {Meier}, \citenamefont {Hegde}, \citenamefont {Ganeshan}, \citenamefont
  {Pixley}, \citenamefont {Vishveshwara},\ and\ \citenamefont
  {Gadway}}]{PhysRevLett.126.040603}%
  \BibitemOpen
  \bibfield  {author} {\bibinfo {author} {\bibfnamefont {F.~A.}\ \bibnamefont
  {An}}, \bibinfo {author} {\bibfnamefont {K.}~\bibnamefont
  {Padavi\ifmmode~\acute{c}\else \'{c}\fi{}}}, \bibinfo {author} {\bibfnamefont
  {E.~J.}\ \bibnamefont {Meier}}, \bibinfo {author} {\bibfnamefont
  {S.}~\bibnamefont {Hegde}}, \bibinfo {author} {\bibfnamefont
  {S.}~\bibnamefont {Ganeshan}}, \bibinfo {author} {\bibfnamefont {J.~H.}\
  \bibnamefont {Pixley}}, \bibinfo {author} {\bibfnamefont {S.}~\bibnamefont
  {Vishveshwara}}, \ and\ \bibinfo {author} {\bibfnamefont {B.}~\bibnamefont
  {Gadway}},\ }\href {\doibase 10.1103/PhysRevLett.126.040603} {\bibfield
  {journal} {\bibinfo  {journal} {Phys. Rev. Lett.}\ }\textbf {\bibinfo
  {volume} {126}},\ \bibinfo {pages} {040603} (\bibinfo {year}
  {2021})}\BibitemShut {NoStop}%
\bibitem [{\citenamefont {Kohlert}\ \emph {et~al.}(2019)\citenamefont
  {Kohlert}, \citenamefont {Scherg}, \citenamefont {Li}, \citenamefont
  {L{\"{u}}schen}, \citenamefont {{Das Sarma}}, \citenamefont {Bloch},\ and\
  \citenamefont {Aidelsburger}}]{PhysRevLett.122.170403}%
  \BibitemOpen
  \bibfield  {author} {\bibinfo {author} {\bibfnamefont {T.}~\bibnamefont
  {Kohlert}}, \bibinfo {author} {\bibfnamefont {S.}~\bibnamefont {Scherg}},
  \bibinfo {author} {\bibfnamefont {X.}~\bibnamefont {Li}}, \bibinfo {author}
  {\bibfnamefont {H.~P.}\ \bibnamefont {L{\"{u}}schen}}, \bibinfo {author}
  {\bibfnamefont {S.}~\bibnamefont {{Das Sarma}}}, \bibinfo {author}
  {\bibfnamefont {I.}~\bibnamefont {Bloch}}, \ and\ \bibinfo {author}
  {\bibfnamefont {M.}~\bibnamefont {Aidelsburger}},\ }\href {\doibase
  10.1103/PhysRevLett.122.170403} {\bibfield  {journal} {\bibinfo  {journal}
  {Phys. Rev. Lett.}\ }\textbf {\bibinfo {volume} {122}},\ \bibinfo {pages}
  {170403} (\bibinfo {year} {2019})}\BibitemShut {NoStop}%
\bibitem [{\citenamefont {Verbin}\ \emph {et~al.}(2015)\citenamefont {Verbin},
  \citenamefont {Zilberberg}, \citenamefont {Lahini}, \citenamefont {Kraus},\
  and\ \citenamefont {Silberberg}}]{PhysRevB.91.064201}%
  \BibitemOpen
  \bibfield  {author} {\bibinfo {author} {\bibfnamefont {M.}~\bibnamefont
  {Verbin}}, \bibinfo {author} {\bibfnamefont {O.}~\bibnamefont {Zilberberg}},
  \bibinfo {author} {\bibfnamefont {Y.}~\bibnamefont {Lahini}}, \bibinfo
  {author} {\bibfnamefont {Y.~E.}\ \bibnamefont {Kraus}}, \ and\ \bibinfo
  {author} {\bibfnamefont {Y.}~\bibnamefont {Silberberg}},\ }\href {\doibase
  10.1103/PhysRevB.91.064201} {\bibfield  {journal} {\bibinfo  {journal} {Phys.
  Rev. B}\ }\textbf {\bibinfo {volume} {91}},\ \bibinfo {pages} {64201}
  (\bibinfo {year} {2015})}\BibitemShut {NoStop}%
\bibitem [{\citenamefont {Sinelnik}\ \emph {et~al.}(2020)\citenamefont
  {Sinelnik}, \citenamefont {Shishkin}, \citenamefont {Yu}, \citenamefont
  {Samusev}, \citenamefont {Belov}, \citenamefont {Limonov}, \citenamefont
  {Ginzburg},\ and\ \citenamefont
  {Rybin}}]{https://doi.org/10.1002/adom.202001170}%
  \BibitemOpen
  \bibfield  {author} {\bibinfo {author} {\bibfnamefont {A.~D.}\ \bibnamefont
  {Sinelnik}}, \bibinfo {author} {\bibfnamefont {I.~I.}\ \bibnamefont
  {Shishkin}}, \bibinfo {author} {\bibfnamefont {X.}~\bibnamefont {Yu}},
  \bibinfo {author} {\bibfnamefont {K.~B.}\ \bibnamefont {Samusev}}, \bibinfo
  {author} {\bibfnamefont {P.~A.}\ \bibnamefont {Belov}}, \bibinfo {author}
  {\bibfnamefont {M.~F.}\ \bibnamefont {Limonov}}, \bibinfo {author}
  {\bibfnamefont {P.}~\bibnamefont {Ginzburg}}, \ and\ \bibinfo {author}
  {\bibfnamefont {M.~V.}\ \bibnamefont {Rybin}},\ }\href {\doibase
  https://doi.org/10.1002/adom.202001170} {\bibfield  {journal} {\bibinfo
  {journal} {Advanced Optical Materials}\ }\textbf {\bibinfo {volume} {8}},\
  \bibinfo {pages} {2001170} (\bibinfo {year} {2020})},\ \Eprint
  {http://arxiv.org/abs/https://onlinelibrary.wiley.com/doi/pdf/10.1002/adom.202001170}
  {https://onlinelibrary.wiley.com/doi/pdf/10.1002/adom.202001170} \BibitemShut
  {NoStop}%
\bibitem [{\citenamefont {Wang}\ \emph
  {et~al.}(2022{\natexlab{a}})\citenamefont {Wang}, \citenamefont {Fu},
  \citenamefont {Peng}, \citenamefont {Kartashov}, \citenamefont {Torner},
  \citenamefont {Konotop},\ and\ \citenamefont {Ye}}]{Wang2022}%
  \BibitemOpen
  \bibfield  {author} {\bibinfo {author} {\bibfnamefont {P.}~\bibnamefont
  {Wang}}, \bibinfo {author} {\bibfnamefont {Q.}~\bibnamefont {Fu}}, \bibinfo
  {author} {\bibfnamefont {R.}~\bibnamefont {Peng}}, \bibinfo {author}
  {\bibfnamefont {Y.~V.}\ \bibnamefont {Kartashov}}, \bibinfo {author}
  {\bibfnamefont {L.}~\bibnamefont {Torner}}, \bibinfo {author} {\bibfnamefont
  {V.~V.}\ \bibnamefont {Konotop}}, \ and\ \bibinfo {author} {\bibfnamefont
  {F.}~\bibnamefont {Ye}},\ }\href {\doibase 10.1038/s41467-022-34394-3}
  {\bibfield  {journal} {\bibinfo  {journal} {Nature Communications}\ }\textbf
  {\bibinfo {volume} {13}},\ \bibinfo {pages} {6738} (\bibinfo {year}
  {2022}{\natexlab{a}})}\BibitemShut {NoStop}%
\bibitem [{\citenamefont {Goblot}\ \emph {et~al.}(2020)\citenamefont {Goblot},
  \citenamefont {{\v{S}}trkalj}, \citenamefont {Pernet}, \citenamefont {Lado},
  \citenamefont {Dorow}, \citenamefont {Lema{\^i}tre}, \citenamefont
  {Le~Gratiet}, \citenamefont {Harouri}, \citenamefont {Sagnes}, \citenamefont
  {Ravets}, \citenamefont {Amo}, \citenamefont {Bloch},\ and\ \citenamefont
  {Zilberberg}}]{Goblot2020Nature}%
  \BibitemOpen
  \bibfield  {author} {\bibinfo {author} {\bibfnamefont {V.}~\bibnamefont
  {Goblot}}, \bibinfo {author} {\bibfnamefont {A.}~\bibnamefont
  {{\v{S}}trkalj}}, \bibinfo {author} {\bibfnamefont {N.}~\bibnamefont
  {Pernet}}, \bibinfo {author} {\bibfnamefont {J.~L.}\ \bibnamefont {Lado}},
  \bibinfo {author} {\bibfnamefont {C.}~\bibnamefont {Dorow}}, \bibinfo
  {author} {\bibfnamefont {A.}~\bibnamefont {Lema{\^i}tre}}, \bibinfo {author}
  {\bibfnamefont {L.}~\bibnamefont {Le~Gratiet}}, \bibinfo {author}
  {\bibfnamefont {A.}~\bibnamefont {Harouri}}, \bibinfo {author} {\bibfnamefont
  {I.}~\bibnamefont {Sagnes}}, \bibinfo {author} {\bibfnamefont
  {S.}~\bibnamefont {Ravets}}, \bibinfo {author} {\bibfnamefont
  {A.}~\bibnamefont {Amo}}, \bibinfo {author} {\bibfnamefont {J.}~\bibnamefont
  {Bloch}}, \ and\ \bibinfo {author} {\bibfnamefont {O.}~\bibnamefont
  {Zilberberg}},\ }\href {\doibase 10.1038/s41567-020-0908-7} {\bibfield
  {journal} {\bibinfo  {journal} {Nature Physics}\ }\textbf {\bibinfo {volume}
  {16}},\ \bibinfo {pages} {832} (\bibinfo {year} {2020})}\BibitemShut
  {NoStop}%
\bibitem [{\citenamefont {Apigo}\ \emph {et~al.}(2019)\citenamefont {Apigo},
  \citenamefont {Cheng}, \citenamefont {Dobiszewski}, \citenamefont {Prodan},\
  and\ \citenamefont {Prodan}}]{PhysRevLett.122.095501}%
  \BibitemOpen
  \bibfield  {author} {\bibinfo {author} {\bibfnamefont {D.~J.}\ \bibnamefont
  {Apigo}}, \bibinfo {author} {\bibfnamefont {W.}~\bibnamefont {Cheng}},
  \bibinfo {author} {\bibfnamefont {K.~F.}\ \bibnamefont {Dobiszewski}},
  \bibinfo {author} {\bibfnamefont {E.}~\bibnamefont {Prodan}}, \ and\ \bibinfo
  {author} {\bibfnamefont {C.}~\bibnamefont {Prodan}},\ }\href {\doibase
  10.1103/PhysRevLett.122.095501} {\bibfield  {journal} {\bibinfo  {journal}
  {Phys. Rev. Lett.}\ }\textbf {\bibinfo {volume} {122}},\ \bibinfo {pages}
  {095501} (\bibinfo {year} {2019})}\BibitemShut {NoStop}%
\bibitem [{\citenamefont {Ni}\ \emph {et~al.}(2019)\citenamefont {Ni},
  \citenamefont {Chen}, \citenamefont {Weiner}, \citenamefont {Apigo},
  \citenamefont {Prodan}, \citenamefont {Al{\`{u}}}, \citenamefont {Prodan},\
  and\ \citenamefont {Khanikaev}}]{Ni2019}%
  \BibitemOpen
  \bibfield  {author} {\bibinfo {author} {\bibfnamefont {X.}~\bibnamefont
  {Ni}}, \bibinfo {author} {\bibfnamefont {K.}~\bibnamefont {Chen}}, \bibinfo
  {author} {\bibfnamefont {M.}~\bibnamefont {Weiner}}, \bibinfo {author}
  {\bibfnamefont {D.~J.}\ \bibnamefont {Apigo}}, \bibinfo {author}
  {\bibfnamefont {C.}~\bibnamefont {Prodan}}, \bibinfo {author} {\bibfnamefont
  {A.}~\bibnamefont {Al{\`{u}}}}, \bibinfo {author} {\bibfnamefont
  {E.}~\bibnamefont {Prodan}}, \ and\ \bibinfo {author} {\bibfnamefont {A.~B.}\
  \bibnamefont {Khanikaev}},\ }\href {\doibase 10.1038/s42005-019-0151-7}
  {\bibfield  {journal} {\bibinfo  {journal} {Communications Physics}\ }\textbf
  {\bibinfo {volume} {2}},\ \bibinfo {pages} {55} (\bibinfo {year}
  {2019})}\BibitemShut {NoStop}%
\bibitem [{\citenamefont {Cheng}\ \emph {et~al.}(2020)\citenamefont {Cheng},
  \citenamefont {Prodan},\ and\ \citenamefont
  {Prodan}}]{PhysRevLett.125.224301}%
  \BibitemOpen
  \bibfield  {author} {\bibinfo {author} {\bibfnamefont {W.}~\bibnamefont
  {Cheng}}, \bibinfo {author} {\bibfnamefont {E.}~\bibnamefont {Prodan}}, \
  and\ \bibinfo {author} {\bibfnamefont {C.}~\bibnamefont {Prodan}},\ }\href
  {\doibase 10.1103/PhysRevLett.125.224301} {\bibfield  {journal} {\bibinfo
  {journal} {Phys. Rev. Lett.}\ }\textbf {\bibinfo {volume} {125}},\ \bibinfo
  {pages} {224301} (\bibinfo {year} {2020})}\BibitemShut {NoStop}%
\bibitem [{\citenamefont {Xia}\ \emph {et~al.}(2020)\citenamefont {Xia},
  \citenamefont {Erturk},\ and\ \citenamefont
  {Ruzzene}}]{PhysRevApplied.13.014023}%
  \BibitemOpen
  \bibfield  {author} {\bibinfo {author} {\bibfnamefont {Y.}~\bibnamefont
  {Xia}}, \bibinfo {author} {\bibfnamefont {A.}~\bibnamefont {Erturk}}, \ and\
  \bibinfo {author} {\bibfnamefont {M.}~\bibnamefont {Ruzzene}},\ }\href
  {\doibase 10.1103/PhysRevApplied.13.014023} {\bibfield  {journal} {\bibinfo
  {journal} {Phys. Rev. Applied}\ }\textbf {\bibinfo {volume} {13}},\ \bibinfo
  {pages} {014023} (\bibinfo {year} {2020})}\BibitemShut {NoStop}%
\bibitem [{\citenamefont {Chen}\ \emph {et~al.}(2021)\citenamefont {Chen},
  \citenamefont {Zhu}, \citenamefont {Tan}, \citenamefont {Wang},\ and\
  \citenamefont {Ma}}]{PhysRevX.11.011016}%
  \BibitemOpen
  \bibfield  {author} {\bibinfo {author} {\bibfnamefont {Z.-G.}\ \bibnamefont
  {Chen}}, \bibinfo {author} {\bibfnamefont {W.}~\bibnamefont {Zhu}}, \bibinfo
  {author} {\bibfnamefont {Y.}~\bibnamefont {Tan}}, \bibinfo {author}
  {\bibfnamefont {L.}~\bibnamefont {Wang}}, \ and\ \bibinfo {author}
  {\bibfnamefont {G.}~\bibnamefont {Ma}},\ }\href {\doibase
  10.1103/PhysRevX.11.011016} {\bibfield  {journal} {\bibinfo  {journal} {Phys.
  Rev. X}\ }\textbf {\bibinfo {volume} {11}},\ \bibinfo {pages} {011016}
  (\bibinfo {year} {2021})}\BibitemShut {NoStop}%
\bibitem [{\citenamefont {Gei}\ \emph {et~al.}(2020)\citenamefont {Gei},
  \citenamefont {Chen}, \citenamefont {Bosi},\ and\ \citenamefont
  {Morini}}]{doi:10.1063/5.0013528}%
  \BibitemOpen
  \bibfield  {author} {\bibinfo {author} {\bibfnamefont {M.}~\bibnamefont
  {Gei}}, \bibinfo {author} {\bibfnamefont {Z.}~\bibnamefont {Chen}}, \bibinfo
  {author} {\bibfnamefont {F.}~\bibnamefont {Bosi}}, \ and\ \bibinfo {author}
  {\bibfnamefont {L.}~\bibnamefont {Morini}},\ }\href {\doibase
  10.1063/5.0013528} {\bibfield  {journal} {\bibinfo  {journal} {Applied
  Physics Letters}\ }\textbf {\bibinfo {volume} {116}},\ \bibinfo {pages}
  {241903} (\bibinfo {year} {2020})},\ \Eprint
  {http://arxiv.org/abs/https://doi.org/10.1063/5.0013528}
  {https://doi.org/10.1063/5.0013528} \BibitemShut {NoStop}%
\bibitem [{\citenamefont {Balents}\ \emph {et~al.}(2020)\citenamefont
  {Balents}, \citenamefont {Dean}, \citenamefont {Efetov},\ and\ \citenamefont
  {Young}}]{Balents2020}%
  \BibitemOpen
  \bibfield  {author} {\bibinfo {author} {\bibfnamefont {L.}~\bibnamefont
  {Balents}}, \bibinfo {author} {\bibfnamefont {C.~R.}\ \bibnamefont {Dean}},
  \bibinfo {author} {\bibfnamefont {D.~K.}\ \bibnamefont {Efetov}}, \ and\
  \bibinfo {author} {\bibfnamefont {A.~F.}\ \bibnamefont {Young}},\ }\href
  {\doibase 10.1038/s41567-020-0906-9} {\bibfield  {journal} {\bibinfo
  {journal} {Nat. Phys.}\ }\textbf {\bibinfo {volume} {16}},\ \bibinfo {pages}
  {725} (\bibinfo {year} {2020})}\BibitemShut {NoStop}%
\bibitem [{\citenamefont {\ifmmode \check{C}\else
  \v{C}\fi{}ade\ifmmode~\check{z}\else \v{z}\fi{}}\ \emph
  {et~al.}(2017)\citenamefont {\ifmmode \check{C}\else
  \v{C}\fi{}ade\ifmmode~\check{z}\else \v{z}\fi{}}, \citenamefont {Mondaini},\
  and\ \citenamefont {Sacramento}}]{PhysRevB.96.144301}%
  \BibitemOpen
  \bibfield  {author} {\bibinfo {author} {\bibfnamefont {T.}~\bibnamefont
  {\ifmmode \check{C}\else \v{C}\fi{}ade\ifmmode~\check{z}\else \v{z}\fi{}}},
  \bibinfo {author} {\bibfnamefont {R.}~\bibnamefont {Mondaini}}, \ and\
  \bibinfo {author} {\bibfnamefont {P.~D.}\ \bibnamefont {Sacramento}},\ }\href
  {\doibase 10.1103/PhysRevB.96.144301} {\bibfield  {journal} {\bibinfo
  {journal} {Phys. Rev. B}\ }\textbf {\bibinfo {volume} {96}},\ \bibinfo
  {pages} {144301} (\bibinfo {year} {2017})}\BibitemShut {NoStop}%
\bibitem [{\citenamefont {Roy}\ \emph {et~al.}(2018)\citenamefont {Roy},
  \citenamefont {Khaymovich}, \citenamefont {Das},\ and\ \citenamefont
  {Moessner}}]{10.21468/SciPostPhys.4.5.025}%
  \BibitemOpen
  \bibfield  {author} {\bibinfo {author} {\bibfnamefont {S.}~\bibnamefont
  {Roy}}, \bibinfo {author} {\bibfnamefont {I.~M.}\ \bibnamefont {Khaymovich}},
  \bibinfo {author} {\bibfnamefont {A.}~\bibnamefont {Das}}, \ and\ \bibinfo
  {author} {\bibfnamefont {R.}~\bibnamefont {Moessner}},\ }\href {\doibase
  10.21468/SciPostPhys.4.5.025} {\bibfield  {journal} {\bibinfo  {journal}
  {SciPost Phys.}\ }\textbf {\bibinfo {volume} {4}},\ \bibinfo {pages} {025}
  (\bibinfo {year} {2018})}\BibitemShut {NoStop}%
\bibitem [{\citenamefont {\ifmmode \check{C}\else
  \v{C}\fi{}ade\ifmmode~\check{z}\else \v{z}\fi{}}\ \emph
  {et~al.}(2019)\citenamefont {\ifmmode \check{C}\else
  \v{C}\fi{}ade\ifmmode~\check{z}\else \v{z}\fi{}}, \citenamefont {Mondaini},\
  and\ \citenamefont {Sacramento}}]{CadeZ2019}%
  \BibitemOpen
  \bibfield  {author} {\bibinfo {author} {\bibfnamefont {T.}~\bibnamefont
  {\ifmmode \check{C}\else \v{C}\fi{}ade\ifmmode~\check{z}\else \v{z}\fi{}}},
  \bibinfo {author} {\bibfnamefont {R.}~\bibnamefont {Mondaini}}, \ and\
  \bibinfo {author} {\bibfnamefont {P.~D.}\ \bibnamefont {Sacramento}},\ }\href
  {\doibase 10.1103/PhysRevB.99.014301} {\bibfield  {journal} {\bibinfo
  {journal} {Phys. Rev. B}\ }\textbf {\bibinfo {volume} {99}},\ \bibinfo
  {pages} {014301} (\bibinfo {year} {2019})}\BibitemShut {NoStop}%
\bibitem [{\citenamefont {Bordia}\ \emph
  {et~al.}(2017{\natexlab{b}})\citenamefont {Bordia}, \citenamefont
  {L{\"u}schen}, \citenamefont {Schneider}, \citenamefont {Knap},\ and\
  \citenamefont {Bloch}}]{Bordia2017}%
  \BibitemOpen
  \bibfield  {author} {\bibinfo {author} {\bibfnamefont {P.}~\bibnamefont
  {Bordia}}, \bibinfo {author} {\bibfnamefont {H.}~\bibnamefont {L{\"u}schen}},
  \bibinfo {author} {\bibfnamefont {U.}~\bibnamefont {Schneider}}, \bibinfo
  {author} {\bibfnamefont {M.}~\bibnamefont {Knap}}, \ and\ \bibinfo {author}
  {\bibfnamefont {I.}~\bibnamefont {Bloch}},\ }\href {\doibase
  10.1038/nphys4020} {\bibfield  {journal} {\bibinfo  {journal} {Nature
  Physics}\ }\textbf {\bibinfo {volume} {13}},\ \bibinfo {pages} {460}
  (\bibinfo {year} {2017}{\natexlab{b}})}\BibitemShut {NoStop}%
\bibitem [{\citenamefont {Sarkar}\ \emph {et~al.}(2021)\citenamefont {Sarkar},
  \citenamefont {Ghosh}, \citenamefont {Sen},\ and\ \citenamefont
  {Sengupta}}]{PhysRevB.103.184309}%
  \BibitemOpen
  \bibfield  {author} {\bibinfo {author} {\bibfnamefont {M.}~\bibnamefont
  {Sarkar}}, \bibinfo {author} {\bibfnamefont {R.}~\bibnamefont {Ghosh}},
  \bibinfo {author} {\bibfnamefont {A.}~\bibnamefont {Sen}}, \ and\ \bibinfo
  {author} {\bibfnamefont {K.}~\bibnamefont {Sengupta}},\ }\href {\doibase
  10.1103/PhysRevB.103.184309} {\bibfield  {journal} {\bibinfo  {journal}
  {Phys. Rev. B}\ }\textbf {\bibinfo {volume} {103}},\ \bibinfo {pages}
  {184309} (\bibinfo {year} {2021})}\BibitemShut {NoStop}%
\bibitem [{\citenamefont {Dumitrescu}\ \emph {et~al.}(2022)\citenamefont
  {Dumitrescu}, \citenamefont {Bohnet}, \citenamefont {Gaebler}, \citenamefont
  {Hankin}, \citenamefont {Hayes}, \citenamefont {Kumar}, \citenamefont
  {Neyenhuis}, \citenamefont {Vasseur},\ and\ \citenamefont
  {Potter}}]{Dumitrescu2022}%
  \BibitemOpen
  \bibfield  {author} {\bibinfo {author} {\bibfnamefont {P.~T.}\ \bibnamefont
  {Dumitrescu}}, \bibinfo {author} {\bibfnamefont {J.~G.}\ \bibnamefont
  {Bohnet}}, \bibinfo {author} {\bibfnamefont {J.~P.}\ \bibnamefont {Gaebler}},
  \bibinfo {author} {\bibfnamefont {A.}~\bibnamefont {Hankin}}, \bibinfo
  {author} {\bibfnamefont {D.}~\bibnamefont {Hayes}}, \bibinfo {author}
  {\bibfnamefont {A.}~\bibnamefont {Kumar}}, \bibinfo {author} {\bibfnamefont
  {B.}~\bibnamefont {Neyenhuis}}, \bibinfo {author} {\bibfnamefont
  {R.}~\bibnamefont {Vasseur}}, \ and\ \bibinfo {author} {\bibfnamefont
  {A.~C.}\ \bibnamefont {Potter}},\ }\href {\doibase
  10.1038/s41586-022-04853-4} {\bibfield  {journal} {\bibinfo  {journal}
  {Nature}\ }\textbf {\bibinfo {volume} {607}},\ \bibinfo {pages} {463}
  (\bibinfo {year} {2022})}\BibitemShut {NoStop}%
\bibitem [{\citenamefont {Shimasaki}\ \emph {et~al.}(2022)\citenamefont
  {Shimasaki}, \citenamefont {Prichard}, \citenamefont {Kondakci},
  \citenamefont {Pagett}, \citenamefont {Bai}, \citenamefont {Dotti},
  \citenamefont {Cao}, \citenamefont {Lu}, \citenamefont {Grover},\ and\
  \citenamefont {Weld}}]{kicked_QC_2022}%
  \BibitemOpen
  \bibfield  {author} {\bibinfo {author} {\bibfnamefont {T.}~\bibnamefont
  {Shimasaki}}, \bibinfo {author} {\bibfnamefont {M.}~\bibnamefont {Prichard}},
  \bibinfo {author} {\bibfnamefont {H.~E.}\ \bibnamefont {Kondakci}}, \bibinfo
  {author} {\bibfnamefont {J.}~\bibnamefont {Pagett}}, \bibinfo {author}
  {\bibfnamefont {Y.}~\bibnamefont {Bai}}, \bibinfo {author} {\bibfnamefont
  {P.}~\bibnamefont {Dotti}}, \bibinfo {author} {\bibfnamefont
  {A.}~\bibnamefont {Cao}}, \bibinfo {author} {\bibfnamefont {T.-C.}\
  \bibnamefont {Lu}}, \bibinfo {author} {\bibfnamefont {T.}~\bibnamefont
  {Grover}}, \ and\ \bibinfo {author} {\bibfnamefont {D.~M.}\ \bibnamefont
  {Weld}},\ }\href {\doibase 10.48550/ARXIV.2203.09442} {\enquote {\bibinfo
  {title} {Anomalous localization and multifractality in a kicked
  quasicrystal},}\ } (\bibinfo {year} {2022})\BibitemShut {NoStop}%
\bibitem [{\citenamefont {Jiang}\ \emph {et~al.}(2019)\citenamefont {Jiang},
  \citenamefont {Lang}, \citenamefont {Yang}, \citenamefont {Zhu},\ and\
  \citenamefont {Chen}}]{PhysRevB.100.054301}%
  \BibitemOpen
  \bibfield  {author} {\bibinfo {author} {\bibfnamefont {H.}~\bibnamefont
  {Jiang}}, \bibinfo {author} {\bibfnamefont {L.-J.}\ \bibnamefont {Lang}},
  \bibinfo {author} {\bibfnamefont {C.}~\bibnamefont {Yang}}, \bibinfo {author}
  {\bibfnamefont {S.-L.}\ \bibnamefont {Zhu}}, \ and\ \bibinfo {author}
  {\bibfnamefont {S.}~\bibnamefont {Chen}},\ }\href {\doibase
  10.1103/PhysRevB.100.054301} {\bibfield  {journal} {\bibinfo  {journal}
  {Phys. Rev. B}\ }\textbf {\bibinfo {volume} {100}},\ \bibinfo {pages}
  {054301} (\bibinfo {year} {2019})}\BibitemShut {NoStop}%
\bibitem [{\citenamefont {Longhi}(2019)}]{PhysRevLett.122.237601}%
  \BibitemOpen
  \bibfield  {author} {\bibinfo {author} {\bibfnamefont {S.}~\bibnamefont
  {Longhi}},\ }\href {\doibase 10.1103/PhysRevLett.122.237601} {\bibfield
  {journal} {\bibinfo  {journal} {Phys. Rev. Lett.}\ }\textbf {\bibinfo
  {volume} {122}},\ \bibinfo {pages} {237601} (\bibinfo {year}
  {2019})}\BibitemShut {NoStop}%
\bibitem [{\citenamefont {Kawabata}\ \emph {et~al.}(2019)\citenamefont
  {Kawabata}, \citenamefont {Shiozaki}, \citenamefont {Ueda},\ and\
  \citenamefont {Sato}}]{PhysRevX.9.041015}%
  \BibitemOpen
  \bibfield  {author} {\bibinfo {author} {\bibfnamefont {K.}~\bibnamefont
  {Kawabata}}, \bibinfo {author} {\bibfnamefont {K.}~\bibnamefont {Shiozaki}},
  \bibinfo {author} {\bibfnamefont {M.}~\bibnamefont {Ueda}}, \ and\ \bibinfo
  {author} {\bibfnamefont {M.}~\bibnamefont {Sato}},\ }\href {\doibase
  10.1103/PhysRevX.9.041015} {\bibfield  {journal} {\bibinfo  {journal} {Phys.
  Rev. X}\ }\textbf {\bibinfo {volume} {9}},\ \bibinfo {pages} {041015}
  (\bibinfo {year} {2019})}\BibitemShut {NoStop}%
\bibitem [{\citenamefont {Liu}\ \emph {et~al.}(2021{\natexlab{a}})\citenamefont
  {Liu}, \citenamefont {Wang}, \citenamefont {Liu}, \citenamefont {Zhou},\ and\
  \citenamefont {Chen}}]{PhysRevB.103.014203}%
  \BibitemOpen
  \bibfield  {author} {\bibinfo {author} {\bibfnamefont {Y.}~\bibnamefont
  {Liu}}, \bibinfo {author} {\bibfnamefont {Y.}~\bibnamefont {Wang}}, \bibinfo
  {author} {\bibfnamefont {X.-J.}\ \bibnamefont {Liu}}, \bibinfo {author}
  {\bibfnamefont {Q.}~\bibnamefont {Zhou}}, \ and\ \bibinfo {author}
  {\bibfnamefont {S.}~\bibnamefont {Chen}},\ }\href {\doibase
  10.1103/PhysRevB.103.014203} {\bibfield  {journal} {\bibinfo  {journal}
  {Phys. Rev. B}\ }\textbf {\bibinfo {volume} {103}},\ \bibinfo {pages}
  {014203} (\bibinfo {year} {2021}{\natexlab{a}})}\BibitemShut {NoStop}%
\bibitem [{\citenamefont {Liu}\ \emph {et~al.}(2021{\natexlab{b}})\citenamefont
  {Liu}, \citenamefont {Zhou},\ and\ \citenamefont
  {Chen}}]{PhysRevB.104.024201}%
  \BibitemOpen
  \bibfield  {author} {\bibinfo {author} {\bibfnamefont {Y.}~\bibnamefont
  {Liu}}, \bibinfo {author} {\bibfnamefont {Q.}~\bibnamefont {Zhou}}, \ and\
  \bibinfo {author} {\bibfnamefont {S.}~\bibnamefont {Chen}},\ }\href {\doibase
  10.1103/PhysRevB.104.024201} {\bibfield  {journal} {\bibinfo  {journal}
  {Phys. Rev. B}\ }\textbf {\bibinfo {volume} {104}},\ \bibinfo {pages}
  {024201} (\bibinfo {year} {2021}{\natexlab{b}})}\BibitemShut {NoStop}%
\bibitem [{\citenamefont {Liu}\ \emph {et~al.}(2021{\natexlab{c}})\citenamefont
  {Liu}, \citenamefont {Wang}, \citenamefont {Zheng},\ and\ \citenamefont
  {Chen}}]{PhysRevB.103.134208}%
  \BibitemOpen
  \bibfield  {author} {\bibinfo {author} {\bibfnamefont {Y.}~\bibnamefont
  {Liu}}, \bibinfo {author} {\bibfnamefont {Y.}~\bibnamefont {Wang}}, \bibinfo
  {author} {\bibfnamefont {Z.}~\bibnamefont {Zheng}}, \ and\ \bibinfo {author}
  {\bibfnamefont {S.}~\bibnamefont {Chen}},\ }\href {\doibase
  10.1103/PhysRevB.103.134208} {\bibfield  {journal} {\bibinfo  {journal}
  {Phys. Rev. B}\ }\textbf {\bibinfo {volume} {103}},\ \bibinfo {pages}
  {134208} (\bibinfo {year} {2021}{\natexlab{c}})}\BibitemShut {NoStop}%
\bibitem [{\citenamefont {Lin}\ \emph {et~al.}(2022)\citenamefont {Lin},
  \citenamefont {Li}, \citenamefont {Xiao}, \citenamefont {Wang}, \citenamefont
  {Yi},\ and\ \citenamefont {Xue}}]{PhysRevLett.129.113601}%
  \BibitemOpen
  \bibfield  {author} {\bibinfo {author} {\bibfnamefont {Q.}~\bibnamefont
  {Lin}}, \bibinfo {author} {\bibfnamefont {T.}~\bibnamefont {Li}}, \bibinfo
  {author} {\bibfnamefont {L.}~\bibnamefont {Xiao}}, \bibinfo {author}
  {\bibfnamefont {K.}~\bibnamefont {Wang}}, \bibinfo {author} {\bibfnamefont
  {W.}~\bibnamefont {Yi}}, \ and\ \bibinfo {author} {\bibfnamefont
  {P.}~\bibnamefont {Xue}},\ }\href {\doibase 10.1103/PhysRevLett.129.113601}
  {\bibfield  {journal} {\bibinfo  {journal} {Phys. Rev. Lett.}\ }\textbf
  {\bibinfo {volume} {129}},\ \bibinfo {pages} {113601} (\bibinfo {year}
  {2022})}\BibitemShut {NoStop}%
\bibitem [{\citenamefont {DeGottardi}\ \emph {et~al.}(2013)\citenamefont
  {DeGottardi}, \citenamefont {Sen},\ and\ \citenamefont
  {Vishveshwara}}]{PhysRevLett.110.146404}%
  \BibitemOpen
  \bibfield  {author} {\bibinfo {author} {\bibfnamefont {W.}~\bibnamefont
  {DeGottardi}}, \bibinfo {author} {\bibfnamefont {D.}~\bibnamefont {Sen}}, \
  and\ \bibinfo {author} {\bibfnamefont {S.}~\bibnamefont {Vishveshwara}},\
  }\href {\doibase 10.1103/PhysRevLett.110.146404} {\bibfield  {journal}
  {\bibinfo  {journal} {Phys. Rev. Lett.}\ }\textbf {\bibinfo {volume} {110}},\
  \bibinfo {pages} {146404} (\bibinfo {year} {2013})}\BibitemShut {NoStop}%
\bibitem [{\citenamefont {Liu}\ \emph {et~al.}(2015)\citenamefont {Liu},
  \citenamefont {Ghosh},\ and\ \citenamefont {Chong}}]{Liu2015}%
  \BibitemOpen
  \bibfield  {author} {\bibinfo {author} {\bibfnamefont {F.}~\bibnamefont
  {Liu}}, \bibinfo {author} {\bibfnamefont {S.}~\bibnamefont {Ghosh}}, \ and\
  \bibinfo {author} {\bibfnamefont {Y.~D.}\ \bibnamefont {Chong}},\ }\href
  {\doibase 10.1103/PhysRevB.91.014108} {\bibfield  {journal} {\bibinfo
  {journal} {Phys. Rev. B - Condens. Matter Mater. Phys.}\ }\textbf {\bibinfo
  {volume} {91}},\ \bibinfo {pages} {014108} (\bibinfo {year}
  {2015})}\BibitemShut {NoStop}%
\bibitem [{\citenamefont {Wang}\ \emph {et~al.}(2016)\citenamefont {Wang},
  \citenamefont {Liu}, \citenamefont {Xianlong},\ and\ \citenamefont
  {Hu}}]{PhysRevB.93.104504}%
  \BibitemOpen
  \bibfield  {author} {\bibinfo {author} {\bibfnamefont {J.}~\bibnamefont
  {Wang}}, \bibinfo {author} {\bibfnamefont {X.-J.}\ \bibnamefont {Liu}},
  \bibinfo {author} {\bibfnamefont {G.}~\bibnamefont {Xianlong}}, \ and\
  \bibinfo {author} {\bibfnamefont {H.}~\bibnamefont {Hu}},\ }\href {\doibase
  10.1103/PhysRevB.93.104504} {\bibfield  {journal} {\bibinfo  {journal} {Phys.
  Rev. B}\ }\textbf {\bibinfo {volume} {93}},\ \bibinfo {pages} {104504}
  (\bibinfo {year} {2016})}\BibitemShut {NoStop}%
\bibitem [{\citenamefont {Deng}\ \emph {et~al.}(2019)\citenamefont {Deng},
  \citenamefont {Ray}, \citenamefont {Sinha}, \citenamefont {Shlyapnikov},\
  and\ \citenamefont {Santos}}]{PhysRevLett.123.025301}%
  \BibitemOpen
  \bibfield  {author} {\bibinfo {author} {\bibfnamefont {X.}~\bibnamefont
  {Deng}}, \bibinfo {author} {\bibfnamefont {S.}~\bibnamefont {Ray}}, \bibinfo
  {author} {\bibfnamefont {S.}~\bibnamefont {Sinha}}, \bibinfo {author}
  {\bibfnamefont {G.~V.}\ \bibnamefont {Shlyapnikov}}, \ and\ \bibinfo {author}
  {\bibfnamefont {L.}~\bibnamefont {Santos}},\ }\href {\doibase
  10.1103/PhysRevLett.123.025301} {\bibfield  {journal} {\bibinfo  {journal}
  {Phys. Rev. Lett.}\ }\textbf {\bibinfo {volume} {123}},\ \bibinfo {pages}
  {025301} (\bibinfo {year} {2019})}\BibitemShut {NoStop}%
\bibitem [{\citenamefont {Wang}\ \emph
  {et~al.}(2020{\natexlab{b}})\citenamefont {Wang}, \citenamefont {Zhang},
  \citenamefont {Niu}, \citenamefont {Yu},\ and\ \citenamefont
  {Liu}}]{PhysRevLett.125.073204}%
  \BibitemOpen
  \bibfield  {author} {\bibinfo {author} {\bibfnamefont {Y.}~\bibnamefont
  {Wang}}, \bibinfo {author} {\bibfnamefont {L.}~\bibnamefont {Zhang}},
  \bibinfo {author} {\bibfnamefont {S.}~\bibnamefont {Niu}}, \bibinfo {author}
  {\bibfnamefont {D.}~\bibnamefont {Yu}}, \ and\ \bibinfo {author}
  {\bibfnamefont {X.-J.}\ \bibnamefont {Liu}},\ }\href {\doibase
  10.1103/PhysRevLett.125.073204} {\bibfield  {journal} {\bibinfo  {journal}
  {Phys. Rev. Lett.}\ }\textbf {\bibinfo {volume} {125}},\ \bibinfo {pages}
  {073204} (\bibinfo {year} {2020}{\natexlab{b}})}\BibitemShut {NoStop}%
\bibitem [{\citenamefont {Liu}\ \emph {et~al.}(2022)\citenamefont {Liu},
  \citenamefont {Xia}, \citenamefont {Longhi},\ and\ \citenamefont
  {Sanchez-Palencia}}]{anomScipost}%
  \BibitemOpen
  \bibfield  {author} {\bibinfo {author} {\bibfnamefont {T.}~\bibnamefont
  {Liu}}, \bibinfo {author} {\bibfnamefont {X.}~\bibnamefont {Xia}}, \bibinfo
  {author} {\bibfnamefont {S.}~\bibnamefont {Longhi}}, \ and\ \bibinfo {author}
  {\bibfnamefont {L.}~\bibnamefont {Sanchez-Palencia}},\ }\href {\doibase
  10.21468/SciPostPhys.12.1.027} {\bibfield  {journal} {\bibinfo  {journal}
  {SciPost Phys.}\ }\textbf {\bibinfo {volume} {12}},\ \bibinfo {pages} {27}
  (\bibinfo {year} {2022})}\BibitemShut {NoStop}%
\bibitem [{\citenamefont {Fraxanet}\ \emph {et~al.}(2022)\citenamefont
  {Fraxanet}, \citenamefont {Bhattacharya}, \citenamefont {Grass},
  \citenamefont {Lewenstein},\ and\ \citenamefont
  {Dauphin}}]{PhysRevB.106.024204}%
  \BibitemOpen
  \bibfield  {author} {\bibinfo {author} {\bibfnamefont {J.}~\bibnamefont
  {Fraxanet}}, \bibinfo {author} {\bibfnamefont {U.}~\bibnamefont
  {Bhattacharya}}, \bibinfo {author} {\bibfnamefont {T.}~\bibnamefont {Grass}},
  \bibinfo {author} {\bibfnamefont {M.}~\bibnamefont {Lewenstein}}, \ and\
  \bibinfo {author} {\bibfnamefont {A.}~\bibnamefont {Dauphin}},\ }\href
  {\doibase 10.1103/PhysRevB.106.024204} {\bibfield  {journal} {\bibinfo
  {journal} {Phys. Rev. B}\ }\textbf {\bibinfo {volume} {106}},\ \bibinfo
  {pages} {024204} (\bibinfo {year} {2022})}\BibitemShut {NoStop}%
\bibitem [{\citenamefont {Johansson}\ and\ \citenamefont
  {Riklund}(1991)}]{PhysRevB.43.13468}%
  \BibitemOpen
  \bibfield  {author} {\bibinfo {author} {\bibfnamefont {M.}~\bibnamefont
  {Johansson}}\ and\ \bibinfo {author} {\bibfnamefont {R.}~\bibnamefont
  {Riklund}},\ }\href {\doibase 10.1103/PhysRevB.43.13468} {\bibfield
  {journal} {\bibinfo  {journal} {Phys. Rev. B}\ }\textbf {\bibinfo {volume}
  {43}},\ \bibinfo {pages} {13468} (\bibinfo {year} {1991})}\BibitemShut
  {NoStop}%
\bibitem [{\citenamefont {Biddle}\ and\ \citenamefont {{Das
  Sarma}}(2010)}]{PhysRevLett.104.070601}%
  \BibitemOpen
  \bibfield  {author} {\bibinfo {author} {\bibfnamefont {J.}~\bibnamefont
  {Biddle}}\ and\ \bibinfo {author} {\bibfnamefont {S.}~\bibnamefont {{Das
  Sarma}}},\ }\href {\doibase 10.1103/PhysRevLett.104.070601} {\bibfield
  {journal} {\bibinfo  {journal} {Phys. Rev. Lett.}\ }\textbf {\bibinfo
  {volume} {104}},\ \bibinfo {pages} {70601} (\bibinfo {year}
  {2010})}\BibitemShut {NoStop}%
\bibitem [{\citenamefont {Bodyfelt}\ \emph {et~al.}(2014)\citenamefont
  {Bodyfelt}, \citenamefont {Leykam}, \citenamefont {Danieli}, \citenamefont
  {Yu},\ and\ \citenamefont {Flach}}]{PhysRevLett.113.236403}%
  \BibitemOpen
  \bibfield  {author} {\bibinfo {author} {\bibfnamefont {J.~D.}\ \bibnamefont
  {Bodyfelt}}, \bibinfo {author} {\bibfnamefont {D.}~\bibnamefont {Leykam}},
  \bibinfo {author} {\bibfnamefont {C.}~\bibnamefont {Danieli}}, \bibinfo
  {author} {\bibfnamefont {X.}~\bibnamefont {Yu}}, \ and\ \bibinfo {author}
  {\bibfnamefont {S.}~\bibnamefont {Flach}},\ }\href {\doibase
  10.1103/PhysRevLett.113.236403} {\bibfield  {journal} {\bibinfo  {journal}
  {Phys. Rev. Lett.}\ }\textbf {\bibinfo {volume} {113}},\ \bibinfo {pages}
  {236403} (\bibinfo {year} {2014})}\BibitemShut {NoStop}%
\bibitem [{\citenamefont {Danieli}\ \emph {et~al.}(2015)\citenamefont
  {Danieli}, \citenamefont {Bodyfelt},\ and\ \citenamefont
  {Flach}}]{PhysRevB.91.235134}%
  \BibitemOpen
  \bibfield  {author} {\bibinfo {author} {\bibfnamefont {C.}~\bibnamefont
  {Danieli}}, \bibinfo {author} {\bibfnamefont {J.~D.}\ \bibnamefont
  {Bodyfelt}}, \ and\ \bibinfo {author} {\bibfnamefont {S.}~\bibnamefont
  {Flach}},\ }\href {\doibase 10.1103/PhysRevB.91.235134} {\bibfield  {journal}
  {\bibinfo  {journal} {Phys. Rev. B}\ }\textbf {\bibinfo {volume} {91}},\
  \bibinfo {pages} {235134} (\bibinfo {year} {2015})}\BibitemShut {NoStop}%
\bibitem [{\citenamefont {Ganeshan}\ \emph {et~al.}(2015)\citenamefont
  {Ganeshan}, \citenamefont {Pixley},\ and\ \citenamefont {{Das
  Sarma}}}]{PhysRevLett.114.146601}%
  \BibitemOpen
  \bibfield  {author} {\bibinfo {author} {\bibfnamefont {S.}~\bibnamefont
  {Ganeshan}}, \bibinfo {author} {\bibfnamefont {J.~H.}\ \bibnamefont
  {Pixley}}, \ and\ \bibinfo {author} {\bibfnamefont {S.}~\bibnamefont {{Das
  Sarma}}},\ }\href {\doibase 10.1103/PhysRevLett.114.146601} {\bibfield
  {journal} {\bibinfo  {journal} {Phys. Rev. Lett.}\ }\textbf {\bibinfo
  {volume} {114}},\ \bibinfo {pages} {146601} (\bibinfo {year}
  {2015})}\BibitemShut {NoStop}%
\bibitem [{\citenamefont {Wang}\ \emph {et~al.}(2021)\citenamefont {Wang},
  \citenamefont {Cheng}, \citenamefont {Liu},\ and\ \citenamefont
  {Yu}}]{PhysRevLett.126.080602}%
  \BibitemOpen
  \bibfield  {author} {\bibinfo {author} {\bibfnamefont {Y.}~\bibnamefont
  {Wang}}, \bibinfo {author} {\bibfnamefont {C.}~\bibnamefont {Cheng}},
  \bibinfo {author} {\bibfnamefont {X.-J.}\ \bibnamefont {Liu}}, \ and\
  \bibinfo {author} {\bibfnamefont {D.}~\bibnamefont {Yu}},\ }\href {\doibase
  10.1103/PhysRevLett.126.080602} {\bibfield  {journal} {\bibinfo  {journal}
  {Phys. Rev. Lett.}\ }\textbf {\bibinfo {volume} {126}},\ \bibinfo {pages}
  {080602} (\bibinfo {year} {2021})}\BibitemShut {NoStop}%
\bibitem [{\citenamefont {Xiao}\ \emph {et~al.}(2021)\citenamefont {Xiao},
  \citenamefont {Xie}, \citenamefont {Dong}, \citenamefont {Chen},
  \citenamefont {Yi},\ and\ \citenamefont {Yan}}]{XIAO20212175}%
  \BibitemOpen
  \bibfield  {author} {\bibinfo {author} {\bibfnamefont {T.}~\bibnamefont
  {Xiao}}, \bibinfo {author} {\bibfnamefont {D.}~\bibnamefont {Xie}}, \bibinfo
  {author} {\bibfnamefont {Z.}~\bibnamefont {Dong}}, \bibinfo {author}
  {\bibfnamefont {T.}~\bibnamefont {Chen}}, \bibinfo {author} {\bibfnamefont
  {W.}~\bibnamefont {Yi}}, \ and\ \bibinfo {author} {\bibfnamefont
  {B.}~\bibnamefont {Yan}},\ }\href {\doibase
  https://doi.org/10.1016/j.scib.2021.07.025} {\bibfield  {journal} {\bibinfo
  {journal} {Science Bulletin}\ }\textbf {\bibinfo {volume} {66}},\ \bibinfo
  {pages} {2175} (\bibinfo {year} {2021})}\BibitemShut {NoStop}%
\bibitem [{\citenamefont {Wang}\ \emph
  {et~al.}(2022{\natexlab{b}})\citenamefont {Wang}, \citenamefont {Zhang},
  \citenamefont {Sun},\ and\ \citenamefont {Liu}}]{mobedges_ext_crit_loc}%
  \BibitemOpen
  \bibfield  {author} {\bibinfo {author} {\bibfnamefont {Y.}~\bibnamefont
  {Wang}}, \bibinfo {author} {\bibfnamefont {L.}~\bibnamefont {Zhang}},
  \bibinfo {author} {\bibfnamefont {W.}~\bibnamefont {Sun}}, \ and\ \bibinfo
  {author} {\bibfnamefont {X.-J.}\ \bibnamefont {Liu}},\ }\href {\doibase
  10.48550/ARXIV.2202.06816} {\enquote {\bibinfo {title} {Quantum phase with
  coexisting localized, extended, and critical zones},}\ } (\bibinfo {year}
  {2022}{\natexlab{b}})\BibitemShut {NoStop}%
\bibitem [{\citenamefont {Azbel}(1979)}]{PhysRevLett.43.1954}%
  \BibitemOpen
  \bibfield  {author} {\bibinfo {author} {\bibfnamefont {M.~Y.}\ \bibnamefont
  {Azbel}},\ }\href {\doibase 10.1103/PhysRevLett.43.1954} {\bibfield
  {journal} {\bibinfo  {journal} {Phys. Rev. Lett.}\ }\textbf {\bibinfo
  {volume} {43}},\ \bibinfo {pages} {1954} (\bibinfo {year}
  {1979})}\BibitemShut {NoStop}%
\bibitem [{\citenamefont {Kohmoto}(1983)}]{PhysRevLett.51.1198}%
  \BibitemOpen
  \bibfield  {author} {\bibinfo {author} {\bibfnamefont {M.}~\bibnamefont
  {Kohmoto}},\ }\href {\doibase 10.1103/PhysRevLett.51.1198} {\bibfield
  {journal} {\bibinfo  {journal} {Phys. Rev. Lett.}\ }\textbf {\bibinfo
  {volume} {51}},\ \bibinfo {pages} {1198} (\bibinfo {year}
  {1983})}\BibitemShut {NoStop}%
\bibitem [{\citenamefont {Gon{\c{c}}alves}\ \emph
  {et~al.}(2022{\natexlab{a}})\citenamefont {Gon{\c{c}}alves}, \citenamefont
  {Amorim}, \citenamefont {Castro},\ and\ \citenamefont
  {Ribeiro}}]{GoncalvesRG2022}%
  \BibitemOpen
  \bibfield  {author} {\bibinfo {author} {\bibfnamefont {M.}~\bibnamefont
  {Gon{\c{c}}alves}}, \bibinfo {author} {\bibfnamefont {B.}~\bibnamefont
  {Amorim}}, \bibinfo {author} {\bibfnamefont {E.~V.}\ \bibnamefont {Castro}},
  \ and\ \bibinfo {author} {\bibfnamefont {P.}~\bibnamefont {Ribeiro}},\ }\href
  {\doibase 10.48550/arxiv.2206.13549} {\  (\bibinfo {year}
  {2022}{\natexlab{a}}),\ 10.48550/arxiv.2206.13549},\ \Eprint
  {http://arxiv.org/abs/2206.13549} {arXiv:2206.13549} \BibitemShut {NoStop}%
\bibitem [{\citenamefont {Aulbach}\ \emph {et~al.}(2004)\citenamefont
  {Aulbach}, \citenamefont {Wobst}, \citenamefont {Ingold}, \citenamefont
  {Hänggi},\ and\ \citenamefont {Varga}}]{Aulbach_2004}%
  \BibitemOpen
  \bibfield  {author} {\bibinfo {author} {\bibfnamefont {C.}~\bibnamefont
  {Aulbach}}, \bibinfo {author} {\bibfnamefont {A.}~\bibnamefont {Wobst}},
  \bibinfo {author} {\bibfnamefont {G.-L.}\ \bibnamefont {Ingold}}, \bibinfo
  {author} {\bibfnamefont {P.}~\bibnamefont {Hänggi}}, \ and\ \bibinfo
  {author} {\bibfnamefont {I.}~\bibnamefont {Varga}},\ }\href {\doibase
  10.1088/1367-2630/6/1/070} {\bibfield  {journal} {\bibinfo  {journal} {New
  Journal of Physics}\ }\textbf {\bibinfo {volume} {6}},\ \bibinfo {pages} {70}
  (\bibinfo {year} {2004})}\BibitemShut {NoStop}%
\bibitem [{SM()}]{SM}%
  \BibitemOpen
  \bibinfo {note} {See supplemental material for: I. Derivation of the
  generalized global duality transformation; II. Details on the calculation of
  local dualities using commensurate approximants; III. Derivation of ratios
  between renormalized couplings; and IV. Multifractal analysis.}\BibitemShut
  {Stop}%
\bibitem [{\citenamefont {Gon{\c{c}}alves}\ \emph
  {et~al.}(2022{\natexlab{b}})\citenamefont {Gon{\c{c}}alves}, \citenamefont
  {Amorim}, \citenamefont {Castro},\ and\ \citenamefont
  {Ribeiro}}]{HdualitiesScipost}%
  \BibitemOpen
  \bibfield  {author} {\bibinfo {author} {\bibfnamefont {M.}~\bibnamefont
  {Gon{\c{c}}alves}}, \bibinfo {author} {\bibfnamefont {B.}~\bibnamefont
  {Amorim}}, \bibinfo {author} {\bibfnamefont {E.~V.}\ \bibnamefont {Castro}},
  \ and\ \bibinfo {author} {\bibfnamefont {P.}~\bibnamefont {Ribeiro}},\ }\href
  {\doibase 10.21468/SciPostPhys.13.3.046} {\bibfield  {journal} {\bibinfo
  {journal} {SciPost Phys.}\ }\textbf {\bibinfo {volume} {13}},\ \bibinfo
  {pages} {046} (\bibinfo {year} {2022}{\natexlab{b}})}\BibitemShut {NoStop}%
\bibitem [{Note1()}]{Note1}%
  \BibitemOpen
  \bibinfo {note} {The local duality function may be very sensitive to the
  choice of the correct dual points. Since their computation was done
  numerically, there is an associated error. Therefore we slightly varied the
  computed dual point and checked that some features may arise only due to a
  slightly incorrect choice of this point (see \protect \citep
  {SM}).}\BibitemShut {Stop}%
\bibitem [{\citenamefont {Janssen}(2004)}]{Janssen}%
  \BibitemOpen
  \bibfield  {author} {\bibinfo {author} {\bibfnamefont {M.}~\bibnamefont
  {Janssen}},\ }\href {\doibase 10.1142/s021797929400049x} {\bibfield
  {journal} {\bibinfo  {journal} {Int. J. Mod. Phys. B}\ }\textbf {\bibinfo
  {volume} {08}},\ \bibinfo {pages} {943} (\bibinfo {year} {2004})}\BibitemShut
  {NoStop}%
\bibitem [{\citenamefont {Gadway}(2015)}]{PhysRevA.92.043606}%
  \BibitemOpen
  \bibfield  {author} {\bibinfo {author} {\bibfnamefont {B.}~\bibnamefont
  {Gadway}},\ }\href {\doibase 10.1103/PhysRevA.92.043606} {\bibfield
  {journal} {\bibinfo  {journal} {Phys. Rev. A}\ }\textbf {\bibinfo {volume}
  {92}},\ \bibinfo {pages} {043606} (\bibinfo {year} {2015})}\BibitemShut
  {NoStop}%
\bibitem [{\citenamefont {Inc.}()}]{Mathematica}%
  \BibitemOpen
  \bibfield  {author} {\bibinfo {author} {\bibfnamefont {W.~R.}\ \bibnamefont
  {Inc.}},\ }\href {https://www.wolfram.com/mathematica} {\enquote {\bibinfo
  {title} {Mathematica, {V}ersion 12.3.1},}\ }\bibinfo {note} {Champaign, IL,
  2021}\BibitemShut {NoStop}%
\end{thebibliography}%

\onecolumngrid

\newpage

\beginsupplement
\begin{center}
\textbf{\large{}Supplemental Material for: \vspace{0.1cm}
}{\large\par}
\par\end{center}
\begin{center}
\begin{large}
Critical phase dualities in 1D exactly-solvable quasiperiodic models
\end{large}
\end{center}

\vspace{0.3cm}

\tableofcontents{}

\section{Derivation of generalized global duality transformation}

Our starting point is the Schrödinger equation

\begin{equation}
h_{n}u_{n}=\sum_{m}e^{i(\alpha-k)(n-m)}f(|n-m|)u_{m},\label{eq:schrodinger_starting_dualities}
\end{equation}
where $h_{n}=\eta-V\chi_{n}(q,\phi)$, $\eta=E+t+V$, $f(|n-m|)=te^{-p|n-m|}$
and $\chi_{m}(q,\phi)=\frac{\sinh q}{\cosh q-\cos(2\pi\tau m+\phi)}$.
$k$ is a phase twist while $\alpha$ was chosen to be a parameter
of the model. From here on we will absorb the parameter $\alpha$
in the twist $k$, that is $k-\alpha\rightarrow k$. In what follows
we will also use the useful identity

\begin{equation}
\chi_{m}(q,\phi)=\frac{\sinh q}{\cosh q-\cos(2\pi\tau m+\phi)}=\sum_{l=-\infty}^{+\infty}e^{-q|l|}e^{i\left(2\pi\tau m+\phi\right)l}.\label{eq:chi_n}
\end{equation}

In Ref.$\,$\citep{HdualitiesScipost} we have seen that generic duality
transformations for Aubry-André-like systems may be defined as

\begin{equation}
\tilde{u}_{n}=\sum_{m}e^{i2\pi\tau nm}W_{m}u_{m}.\label{eq:dual_transf}
\end{equation}
Here we will find the form of $W_{m}$ for which an exact global duality
of the type in Eq.$\,$\eqref{eq:dual_transf} can be defined. The
inverse transformation is

\begin{equation}
u_{m}=\frac{1}{N}W_{m}^{-1}\sum_{n}e^{-i2\pi\tau nm}\tilde{u}_{n},
\end{equation}
where $N$ is the total number of sites in the system. Writing Eq.$\,$\eqref{eq:schrodinger_starting_dualities}
in terms of the dual wave function $\tilde{u}_{n}$, we have

\begin{equation}
\begin{aligned}h_{n}W_{n}^{-1}\sum_{m}e^{-i2\pi\tau nm}\tilde{u}_{m}=\sum_{m}e^{-ik(n-m)}f(|n-m|)W_{m}^{-1}\sum_{l}e^{-i2\pi\tau ml}\tilde{u}_{l}\end{aligned}
.
\end{equation}
Multiplying by $e^{i2\pi\tau n\mu}$ and summing over $n$, we get

\begin{equation}
\sum_{m}\Bigg[\sum_{n}e^{i2\pi\tau n(\mu-m)}W_{n}^{-1}\Big(h_{n}-\chi_{\mu}(p,-k)\Big)\Bigg]\tilde{u}_{m}=0.\label{eq:dual_eq}
\end{equation}

Our aim now is to find $W_{n}$ such that Eq.$\,$\eqref{eq:dual_eq}
for some point $P(t,V,p,q,\alpha,E;\phi,k)$, is dual of Eq.$\,$\eqref{eq:schrodinger_starting_dualities}
at a dual point $P'(t',V',p',q',\alpha',E';\phi',k')$. For convenience,
we write Eq.$\,$\eqref{eq:schrodinger_starting_dualities} as

\begin{equation}
\Lambda_{\mu}\Big(\sum_{m}[h_{m}\delta_{m,\mu}-e^{-ik(\mu-m)}e^{-p|\mu-m|}]u_{m}\Big)=0.
\end{equation}
Notice that $\Lambda_{\mu}$ is an additional degree of freedom that
does not affect the solution of Eq.$\,$\eqref{eq:schrodinger_starting_dualities}:
we want to inspect when equations \eqref{eq:schrodinger_starting_dualities}
and \eqref{eq:dual_eq} are equal under a suitable choice of $\Lambda_{k}$.
Imposing an equality for these equations term by term, we have

\begin{equation}
\Lambda_{\mu}\Big(h_{m}\delta_{m,\mu}-e^{-ik(\mu-m)}e^{-p|\mu-m|}\Big)=\sum_{n}e^{i2\pi\tau n(\mu-m)}W_{n}^{-1}\Big(h_{n}'-\chi_{\mu}(p',-k')\Big).\label{eq:starting_eq_Wn}
\end{equation}
 From Eq.$\,$\eqref{eq:starting_eq_Wn}, we obtain

\begin{equation}
W_{l}=\gamma_{\mu}\frac{h_{l}'-\chi_{\mu}(p',-k')}{h_{\mu}-\chi_{l}(p,k)}=\gamma_{\mu}\frac{\eta'-V'\chi_{l}(q',\phi')-\chi_{\mu}(p',-k')}{\eta-V\chi_{\mu}(q,\phi)-\chi_{l}(p,k)}\equiv\gamma_{\mu}P_{\mu l},
\end{equation}
where $\gamma_{\mu}=\Lambda_{\mu}^{-1}$. Note that the right-hand-side
is a tensor. This means that $W_{l}$ is only well defined, that is,
a duality transformation of the type in Eq.$\,$\eqref{eq:dual_transf}
only exists, if we can write $P_{\mu l}=p_{\mu}W_{l}$, so that $\gamma_{\mu}$
can be chosen as $p_{\mu}^{-1}$. This cannot be done in general.
Defining $c_{\mu,k}=\cos(2\pi\tau\mu+k)$ and $c_{\mu,\phi}=\cos(2\pi\tau\mu+\phi)$,
we have

\begin{equation}
\begin{aligned}W_{l} & =\gamma_{\mu}\Bigg(\frac{c_{\mu,\phi}-\cosh q}{c_{\mu,-k'}-\cosh p'}\Bigg)\Bigg(\frac{c_{l,k}-\cosh p}{c_{l,\phi'}-\cosh q'}\Bigg)\frac{B(V',\eta',p')}{A(V,\eta,q)}\Bigg(\frac{\frac{D(V',\eta',p',q')}{B(V',\eta',p')}+\frac{A(V',\eta',q')}{B(V',\eta',p')}c_{\mu,-k'}+c_{l,\phi'}+\frac{\eta'}{B(V',\eta',p')}c_{\mu,-k'}c_{l,\phi'}}{\frac{D(V,\eta,p,q)}{A(V,\eta,q)}+\frac{B(V,\eta,p)}{A(V,\eta,q)}c_{\mu,\phi}+c_{l,k}+\frac{\eta}{A(V,\eta,q)}c_{\mu,\phi}c_{l,k}}\Bigg)\end{aligned}
\label{eq:Wl_expression}
\end{equation}
where, as in the main text, we have

\begin{equation}
\begin{aligned}A(V,\eta,q)= & -\eta\cosh q+V\sinh q\\
B(V,\eta,p)= & -\eta\cosh p+\sinh p\\
D(V,\eta,p,q)= & \eta\cosh p\cosh q-\cosh q\sinh p-V\cosh p\sinh q
\end{aligned}
.
\end{equation}

The last term in Eq.$\,$\eqref{eq:Wl_expression} is the only problematic
term, for which the indexes $\mu$ and $l$ do not decouple. However,
if we require the numerator and denominator to be equal, this term
will be equal to $1$ and $W_{l}$ becomes well-defined for any choice
of $\gamma_{\mu}$ that cancels the $\mu$-dependence. This is, for
instance, the case if (note the change of phases $\phi$ and $k$):

\begin{equation}
\begin{cases}
k'=-\phi;{\rm \textrm{ }}\phi'=k\\
\frac{D(V',\eta',p',q')}{B(V',\eta',p')}=\frac{D(V,\eta,p,q)}{A(V,\eta,q)}\\
\frac{A(V',\eta',q')}{B(V',\eta',p')}=\frac{B(V,\eta,p)}{A(V,\eta,q)}\\
\frac{\eta'}{B(V',\eta',p')}=\frac{\eta}{A(V,\eta,q)}
\end{cases}.\label{eq:dual_1}
\end{equation}
The choice

\begin{equation}
\begin{cases}
k'=-\phi+\pi;\textrm{ }\phi'=k+\pi\\
\frac{D(V',\eta',p',q')}{B(V',\eta',p')}=-\frac{D(V,\eta,p,q)}{A(V,\eta,q)}\\
\frac{A(V',\eta',q')}{B(V',\eta',p')}=\frac{B(V,\eta,p)}{A(V,\eta,q)}\\
\frac{\eta'}{B(V',\eta',p')}=-\frac{\eta}{A(V,\eta,q)}
\end{cases}\label{eq:dual_2}
\end{equation}
is equally valid. The first choice implies that $A(V,\eta,q)=B(V,\eta,p)$
at SD points, while the second implies that $A(V,\eta,q)=-B(V,\eta,p)$.
Alternatively, we could choose $k'=-\phi+\pi$ and $\phi'=k$ or $k'=-\phi$
and $\phi'=k+\pi$. But these choices give rise to contradicting equations
at the SD points. The equations for dual points written in Eqs.$\,$\eqref{eq:dual_1}
and~\eqref{eq:dual_2} are summarized in the main text.

Finally, if we choose

\begin{equation}
\gamma_{\mu}=\frac{A(V,\eta,q)\sinh q(c_{\mu,-k'}-\cosh p')}{B(V',\eta',p')\sinh p(c_{\mu,\phi}-\cosh q)},
\end{equation}
we get, from Eq.$\,$\eqref{eq:Wl_expression}, that

\begin{equation}
W_{l}=\chi_{l}(q',\phi')\chi_{l}^{-1}(p,k).\label{eq:Wl_final}
\end{equation}

We finally check some limiting cases that were already known in the
literature. The first case is the Aubry-André model which is recovered
from our model if we make the substitutions $t\rightarrow te^{p}$
and $V\rightarrow Ve^{q}/2$ and then take the large $p$ limit, with
$p=q=p'=q'$. Nonetheless, even if we do not take the large $p$ limit,
just by setting these latter equalities we get $W_{l}=1$, implying
that $\tilde{u}_{n}=\sum_{m}e^{i2\pi\tau nm}u_{m}$, which is just
the original Aubry-André duality transformation \citep{AubryAndre}.
The dual points can also be obtained from Eqs.$\,$\eqref{eq:dual_1},\eqref{eq:dual_2}
to be $V'=4t^{2}/V$ and $E'=\pm2Et/V$, as obtained in Ref.~\citep{AubryAndre}
(the $\pm$ signs come respectively from Eqs.$\,$\eqref{eq:dual_1},\eqref{eq:dual_2}).

We can also check the large $q$ limit that corresponds to the limit
in \citep{PhysRevLett.104.070601} if we make $t\rightarrow te^{p}$
and $V\rightarrow Ve^{q}$. In this case, $\chi_{l}(q',\phi')\rightarrow1$
and we obtain the duality transformation in this reference. On the
other hand, in the large $p$ limit we get the model in Ref.~\citep{PhysRevLett.114.146601}
making again the previous substitutions. In this limit, $\chi_{l}^{-1}(p,k)\rightarrow1$
and we get the dual transformation proven in Ref.~\citep{HdualitiesScipost}
at SD points. Even though the duality transformations were used only
to find SD points in Refs.~\citep{PhysRevLett.104.070601,PhysRevLett.114.146601},
they can also be used to define other dual points that can be found
through Eqs.$\,$\ref{eq:dual_1}, \ref{eq:dual_2}.

We finally remark that away from the previously mentioned limits,
the duality transformation depends both on the starting and dual points.
Even though this was a possibility introduced in Ref.~\citep{HdualitiesScipost},
the transformation in Eq.$\,$\eqref{eq:Wl_final} was, as far as
the authors are aware, the first found global exact duality of this
type.

\section{Details on the calculation of local dualities using commensurate
approximants}

\label{sec:duality_function_details}

To compute the data points used in Fig.$\,$2(b),
corresponding to samples of the duality function $W'(x)$, we followed
the procedure introduced in Ref.~\citep{HdualitiesScipost}. We first
define dual points $P$ and $P'$ respectively at energies $E(\varphi_{0},\kappa_{0},\bm{\lambda})$
and $E\text{'}(\varphi_{0},\kappa_{0},\bm{\lambda}')$, where $\varphi=L\phi$
and $\kappa=Lk$ as in the main text and $\bm{\lambda}$, $\bm{\lambda}'$
denote vectors with the Hamiltonian parameters at $P$ and $P'$.
Using the definition of dual points in Ref.~\citep{HdualitiesScipost},
$P$ and $P'$ are found as being points for which the energy dispersions
$\Delta E_{\kappa}=E(\varphi_{0},\kappa_{0}+\delta\kappa,\bm{\lambda})-E(\varphi_{0},\kappa_{0},\bm{\lambda})$
around $P$ is equal to the energy dispersion $\Delta E'_{\varphi}=E'(\varphi_{0}+\delta\varphi,\kappa_{0},\bm{\lambda}')-E'(\varphi_{0},\kappa_{0},\bm{\lambda}')$,
with small $\delta\varphi$ and $\delta\kappa$. In the following,
we will use single-particle wave functions with $\phi=k=0$ to define
$W'(x)$ and therefore set $\varphi_{0}=\kappa_{0}=0$. For a given
CA defined by $\tau_{c}=L'/L$ ($L,L'$ co-prime integers), we define
$u^{\text{r}}(P')$ as the solution to Eq.$\,$(3) of the main text 
at point $P'$ and $u^{\text{d}}(P)$ as

\begin{equation}
u_{n}^{\t d}(P)=\frac{1}{\sqrt{L}}\sum_{m=0}^{L-1}e^{i2\pi\tau_{c}mn}u_{n}^{\t r}(P).
\end{equation}

As in Ref.~\citep{HdualitiesScipost}, we define the duality matrix
$\mathcal{O}_{c}$ in terms of $u^{\text{r}}(P')$ and $u^{\t d}(P)$
as

\begin{equation}
\mathcal{O}_{c}[T^{n}u^{\t d}(P)]=T^{n}u^{\text{r}}(P'),\hspace{1em}n=0,\cdots,L-1,
\end{equation}
where $T$ is the cyclic translation operator defined as $T\psi=\psi'$
with $\psi'_{i}=\psi_{\mod(i+1,L)}$. Since $\mathcal{O}_{c}$ is
a circulant matrix, we may write it as

\begin{equation}
\mathcal{O}_{c}=U^{\dagger}W'U
\end{equation}
where $U$ is a matrix with entries $U_{\mu\nu}=e^{2\pi i\frac{L'}{L}\mu\nu}$
and $W'$ is a diagonal matrix $W'_{\mu\nu}=w'_{\mu}\delta_{\mu\nu}$
with the eigenvalues $w'_{\mu}$ of $\mathcal{O}_{c}$. We can therefore
write

\begin{equation}
u^{\text{r}}(P')=U^{\dagger}W'u^{\t r}(P)\leftrightarrow u_{\mu}^{\text{r}}(P')=\sum_{\nu=0}^{L-1}e^{2\pi i\tau_{c}\mu\nu}w'_{\nu}u_{\nu}^{\t r}(P).
\end{equation}

The eigenvalues $w'_{\nu}$ are, as seen in Ref.~\citep{HdualitiesScipost},
evaluations of a function $W'(x)$, that has period $\Delta x=1$,
at points $x_{\nu}=\mod(\nu\tau_{c},1),\textrm{ }\nu=0,\cdots,L-1$.
This function is continuously sampled in the limit that $\tau_{c}\rightarrow\tau$,
that is, as $L\rightarrow\infty$. At global dual points defined by
Eq.$\,$5 of the main text, the exact duality
transformation is given by

\begin{equation}
\tilde{u}_{\mu}(P')=\sum_{\nu}e^{i2\pi\tau\mu\nu}W(\tau\nu)u_{\nu}(P)\label{eq:duality_transformation-1}
\end{equation}
where $W(x)=\chi(q',x)\chi^{-1}(p,x)$ and $\chi(\lambda,x)=\sinh\lambda[\cosh\lambda-\cos(2\pi x)]^{-1}$.
In the commensurate limit, replacing $\tau\rightarrow\tau_{c}$ in
Eq.$\,$\eqref{eq:duality_transformation-1}, we find that

\begin{equation}
w'_{\nu}=W'(\tau_{c}\nu)\propto W(\tau_{c}\nu).
\end{equation}

The duality function $W(x)$ can therefore be sampled at $L$ different
points by computing the eigenvalues $w'_{\nu}$ for CA with $L$ sites
in the unit cell. They were computed and shown as data points in Fig.$\,$2(b)
of the main text together with the exact analytical duality function
to confirm the validity of the latter.

In the case of local dualities, the exact duality function is unknown
and the only way to access it is by computing the eigenvalues $w'_{\nu}$
at (locally) dual points. In this case, dual points were computed
numerically as explained in Ref.~\citep{HdualitiesScipost}. As shown
in one of the examples in Fig.$\,$2(b)
of the main text, the duality function for the local dualities can
be highly non-trivial.

\begin{figure}[h]
\centering{}\includegraphics[width=0.85\columnwidth]{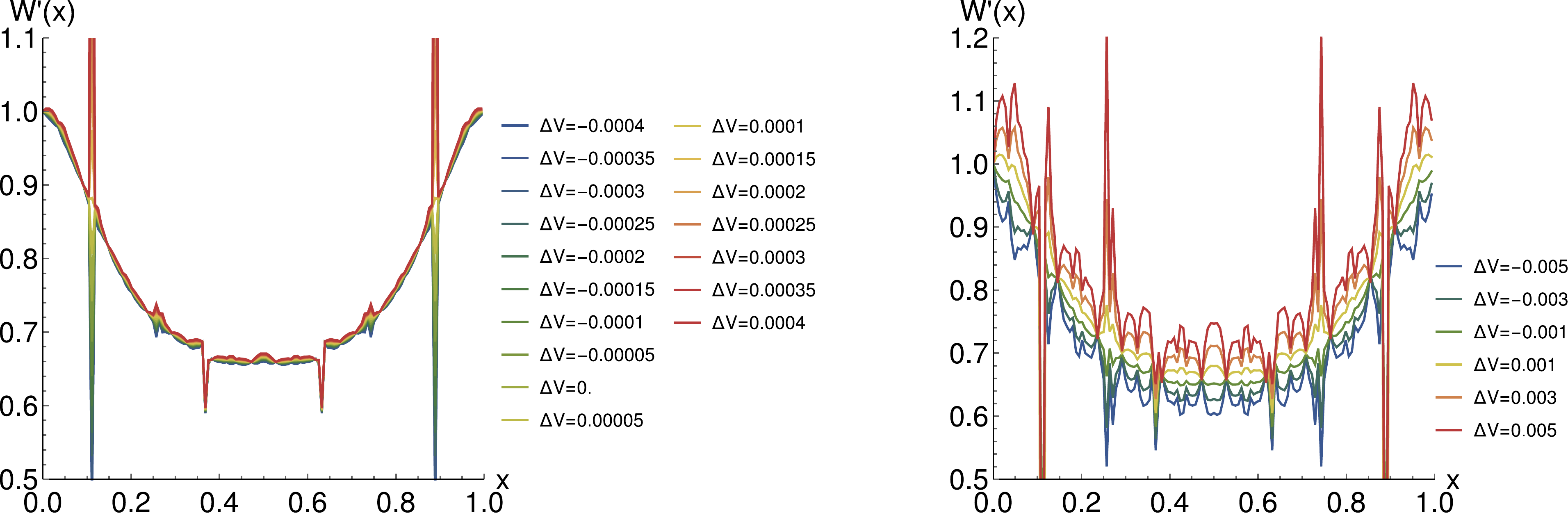}\caption{$W'(x)$ obtained by linear interpolation of the eigenvalues $w'_{\nu}$
for $\tau_{c}=89/144$, for the parameters defining the duality transformation
in Fig.$\,$2(b) (bottom, red) of
the main text. We considered a starting point $P$ with $p=1.3,q=1,V\approx0.73015,E\approx0.34580$
and points $P'$ with $p'=1.3,q'=1,V'\approx0.67439+\Delta V,E'\approx0.33609$.
The point for $\Delta V=0$ corresponds to the dual point computed
numerically. In fact, the smoothest function is obtained for the point
with $\Delta V\approx0.0001$ for which the sharp features close to
$x=0.1$ and $x=0.9$ are removed (left figure). This suggests that
the latter is the true dual point and that even a small numerical
error may lead to the appearance of artificial features. If we move
away from the true dual point (right figure), the duality function
becomes very noisy and meaningless.\label{fig:dual_sensitiveness}}
\end{figure}

We finish by remarking that it is important to precisely compute the
dual points to get a meaningful duality function. While for the global
dualities the dual points can be calculated exactly, this is not the
case for the local dualities for which there are unavoidable numerical
errors. In Fig.$\,$\ref{fig:dual_sensitiveness} we show an example
where it can be seen that even a small numerical error in the calculation
of dual points may introduce artificial features to the function $W'(x)$:
we must therefore slightly change the dual point with respect to the
computed one and check the robustness of function $W'(x)$. In this
example we checked that the dual point was miscalculated with an error
in the parameter $V$ of $\Delta V\approx0.0001$, that was corrected
to present the results of Fig.$\,$2(b)
of the main text.

\section{Derivation of ratios between renormalized couplings}

Our starting point is Eq.$\,$\eqref{eq:schrodinger_starting_dualities}
for a commensurate system defined by $\tau=\tau_{c}=L'/L$, where
$L'$ and $L$ two co-prime integers and $L$ defines the number of
sites in the unit cell. In this case, we have $h_{n+rL}=h_{n}$ and
Bloch's theorem warrants $u_{n+rL}=u_{n},n=0,\cdots,L-1,{\rm }r\in\mathbb{Z}$.

We start by considering the example of one site per unit cell, that
is, $\tau_{c}=1$. In this case, the Schrödinger equation becomes
\[
\left[h_{0}-t\sum_{m}e^{ikm}f(|m|)\right]u_{0}=0,
\]
which can be written as
\[
\left[\eta-V\chi_{0}(q,\phi)-t\chi_{0}(p,k)\right]u_{0}=0,
\]
with $\chi_{0}$ given by Eq.~\eqref{eq:chi_n}. Noticing that $[\cosh q-\cos(\phi)][\cosh p-\cos(k)]\neq0$
we have
\begin{equation}
\left[A(V,\eta,q)\cos(k)+B(V,\eta,p)\cos(\phi)+\eta\cos(\phi)\cos(k)+D(V,\eta,p,q)\right]=0.\label{eq:CA_tauc-1}
\end{equation}

Equation.$\,$\eqref{eq:CA_tauc-1} defines the characteristic polynomial
for the simplest CA. To characterize higher-order CA, we start by
applying the transformation $\sum_{n=0}^{L-1}e^{i2\pi\tau_{c}n\mu}=\frac{L}{N}\sum_{n}e^{i2\pi\tau_{c}n\mu}$,
where $N$ is the total number of sites in the system. The first term
in Eq.$\,$\eqref{eq:schrodinger_starting_dualities} becomes

\begin{equation}
\sum_{n=0}^{L-1}e^{i2\pi\tau_{c}n\mu}h_{n}u_{n}=\frac{L}{N}\sum_{m}e^{i2\pi\tau_{c}m\mu}h_{m}u_{m},{\rm }\textrm{with }h_{m+rL}=h_{m},r\in\mathbb{Z}.\label{eq:hn_basis}
\end{equation}
For the second term we have (absorbing again $\alpha$ in the phase
twist $k$):

\begin{equation}
\begin{alignedat}{1} & \frac{L}{N}\sum_{n}e^{i2\pi\tau_{c}n\mu}\sum_{m}e^{-ik(n-m)}f(|n-m|)u_{m}\\
 & \textrm{Using \ensuremath{m'=n-m}:}\\
 & \frac{L}{N}\sum_{m'}e^{i2\pi\tau_{c}\mu m'}e^{-ikm'}f(|m'|)\sum_{m}e^{i2\pi\tau_{c}m\mu}u_{m}=\frac{L}{N}\chi(p,2\pi\tau_{c}\mu-k)\sum_{m}e^{i2\pi\tau_{c}m\mu}u_{m},
\end{alignedat}
\label{eq:hop_basis}
\end{equation}
where we used Eq.$\,$\eqref{eq:chi_n} and defined $\chi(\lambda,x)\equiv\chi_{0}(\lambda,x)$.
In the commensurate case, we only sample the function $\chi(\lambda,x)$
at a discrete set of points $x_{\mu}=2\pi\tau_{c}\mu-k,{\rm }\mu=0,\cdots,L-1$.
Combining Eqs.$\,$\eqref{eq:hn_basis} and~\eqref{eq:hop_basis},
we get

\begin{equation}
\begin{aligned}\sum_{m}e^{i2\pi\tau_{c}m\mu}[h_{m}-\chi(p,2\pi\tau_{c}\mu-k)]u_{m} & =0\\
\Leftrightarrow\sum_{m}e^{i2\pi\tau_{c}m\mu}\Bigg(\frac{A(V,\eta,q)c_{\mu,-k}+B(V,\eta,p)c_{m,\phi}+\eta c_{m,\phi}c_{\mu,-k}+D(V,\eta,p,q)}{[\cosh q-c_{m,\phi}][\cosh p-c_{\mu,-k}]}\Bigg)u_{m} & =0\\
\Leftrightarrow M\bm{u} & =0
\end{aligned}
\label{eq:M_eq}
\end{equation}
where we used the short-hand notation $c_{m,\phi}=\cos(2\pi\tau_{c}m+\phi)$
and $c_{\mu,k}=\cos(2\pi\tau_{c}\mu+k)$ and $\bm{u}=(u_{0},\cdots,u_{L-1})$.
Note that each component $M_{\mu m}$ is now written in a form that
resembles Eq.$\,$\eqref{eq:CA_tauc-1} for the simplest CA. In particular,
we essentially made appear the characteristic polynomial for this
CA for every term, with the difference that now the phase $\phi$
is associated with index $m$ and the phase $-k$ with index $\mu$.

We will now compute the $\phi$- and $k$-dependent parts of the determinant
of matrix $M$ in Eq.$\,$\eqref{eq:M_eq}. Our final aim is to calculate
the ratios between renormalized couplings $|t_{L}/C_{L}|,|V_{L}/C_{L}|$
and $|t_{L}/V_{L}|$, where these couplings are defined, for the characteristic
polynomial of a CA with $L$ sites in the unit cell, $\mathcal{P}_{L}(\varphi,\kappa)$,
as

\begin{equation}
\begin{aligned}\mathcal{P}_{L}(\varphi,\kappa)= & V_{L}\cos(\varphi)+t_{L}\cos(\kappa)+C_{L}\cos(\varphi)\cos(\kappa)+\cdots\end{aligned}
\end{equation}
where $\varphi=L\phi$ and $\kappa=Lk$. On the way, we will also
explain why only the fundamental harmonics in $\varphi$ and $\kappa$
appear for any CA.

We start by noting that the denominator of each term $M_{\mu m}$
can be written as $T_{\mu\phi}T'_{mk}=[\cosh q-c_{\mu,\phi}][\cosh p-c_{m,-k}]$.
The Leibniz formula for the determinant is

\begin{equation}
{\rm det}(M)=\sum_{\bm{\sigma}\in S_{L}}{\rm sgn}(\sigma)\prod_{\mu=0}^{L-1}M_{\mu,\sigma_{\mu}}
\end{equation}
where $S_{L}$ is the set of permutations of indexes $i=1,\cdots,L$.
Therefore for each term in the sum we get all the possible combinations
of $T_{\mu\phi}T'_{mk}$, that is if we write $M_{\mu m}=P_{\mu m}/(T_{\mu\phi}T'_{mk})$,
we have

\begin{equation}
{\rm det}(M)=\frac{1}{\prod_{\mu}T_{\mu\phi}T'_{\mu k}}\sum_{\bm{\sigma}\in S_{L}}{\rm sgn}(\sigma)\prod_{\mu=0}^{L-1}P_{\mu,\sigma_{\mu}}\propto\sum_{\bm{\sigma}\in S_{L}}{\rm sgn}(\sigma)\prod_{\mu=0}^{L-1}P_{\mu,\sigma_{\mu}}.
\end{equation}

Let us focus on the $|t_{L}/C_{L}|$ ratio. We have

\begin{equation}
{\rm det}(M)\propto\sum_{\bm{\sigma}\in S_{L}}{\rm sgn}(\sigma)\prod_{\mu=0}^{L-1}e^{i2\pi\tau_{c}\mu\sigma_{\mu}}\Bigg(\eta c_{\mu,-k}(\frac{A}{\eta}+c_{\sigma_{\mu},\phi})+Bc_{\sigma_{\mu},\phi}+D\Bigg).\label{eq:detM_2}
\end{equation}
We first realize that only products of terms $\eta c_{\mu,-k}(A/\eta+c_{\sigma_{\mu},\phi})$
matter to get the terms $t_{L}\cos(\kappa)+C_{L}\cos(\kappa)\cos(\varphi)$
of $\mathcal{P}_{L}(\varphi,\kappa)$. This is because to obtain terms
with $L$ times the original frequency $k$ or $\phi$ (recall the
definitions $\varphi=L\phi$ and $\kappa=Lk$), we need to have $L$
products of terms $c_{\mu,-k}$ and $c_{\mu,\phi}$. The only way
to accomplish this is by multiplying $L$ terms of the type $\eta c_{\mu,-k}(A/\eta+c_{\sigma_{\mu},\phi})$.
At this point, we can also ask why terms of the type $\cos(nx)$ with
$x=\phi,k$ and $n\in\mathbb{N}<L$ do not appear in $\mathcal{P}_{L}(\varphi,\kappa)$.
This would give rise to periodicities in $\phi$ and $k$, $\Delta\phi,\Delta k>2\pi/L$,
that are forbidden. The twist $k$ is nothing more than a Bloch momentum
associated with the unit cell of size $L$: the reciprocal lattice
vector for this unit cell is $G=2\pi/L$ and therefore the energy
bands should repeat with a period $\Delta k=2\pi/L$. In the case
of phase $\phi$, shifts $\Delta\phi=2\pi/L$ in a CA with a unit
cell with $L$ sites are just re-definitions of this unit cell \citep{HdualitiesScipost}.
The energy bands should therefore be periodic upon these shifts. Finally,
we can also ask why we cannot have terms of the type $\cos(nx)$ with
$x=\phi,k$ and $n\in\mathbb{N}>L$. The reason is that such terms
would require a number of $n>L$ products of terms $c_{\mu,-k}$ and
$c_{\mu,\phi}$ that do not appear in the determinant in Eq.$\,$\ref{eq:detM_2}.
Therefore, we only have fundamental harmonics in $\varphi$ and $\kappa$,
as previously stated.

To make further progress, we will need the following identity for
$L,L'$ two co-prime integers \citep{Mathematica},
\begin{equation}
\prod_{\mu=0}^{L-1}\Bigg[\cos(y)-\cos\Big(\frac{2\pi L'\mu}{L}+\phi\Big)\Bigg]=2^{1-L}\Big[\cos(Ly)-\cos(L\phi)\Big],
\end{equation}
 which we can apply to the term $\sum_{\bm{\sigma}\in S_{L}}{\rm sgn}(\sigma)\prod_{\mu=0}^{L-1}e^{i2\pi\tau_{c}\mu\sigma_{\mu}}\eta c_{\mu,-k}(\frac{A}{\eta}+c_{\sigma_{\mu},\phi})$
in Eq.$\,$\eqref{eq:detM_2}, as long as $|A/\eta|<1$ (which occurs
inside the critical phase) by identifying $y=\arccos A/\eta$. We
then obtain

\begin{equation}
\begin{aligned}{\rm det}(M)\propto & \gamma_{\bm{\sigma}}\Big(-\cos(L\pi/2)+\cos(Lk)\Big)\Big(\cos[L\arccos(A/\eta)]-\cos(L\phi)\Big)+\sum_{\bm{\sigma}\in S_{L}}{\rm sgn}(\sigma)\prod_{\mu=0}^{L-1}e^{i2\pi\tau_{c}\mu\sigma_{\mu}}\Big(Bc_{\sigma_{\mu},\phi}+D\Big)\\
\propto & \cos[L\arccos(A/\eta)]\cos(Lk)-\cos(Lk)\cos(L\phi)+\cdots
\end{aligned}
\end{equation}
where we identified

\begin{equation}
\gamma_{\bm{\sigma}}=\sum_{\bm{\sigma}\in S_{L}}{\rm sgn}(\sigma)\prod_{\mu=0}^{L-1}e^{i2\pi\tau_{c}\mu\sigma_{\mu}}.
\end{equation}

\begin{figure}[h]
\begin{centering}
\includegraphics[width=0.75\columnwidth]{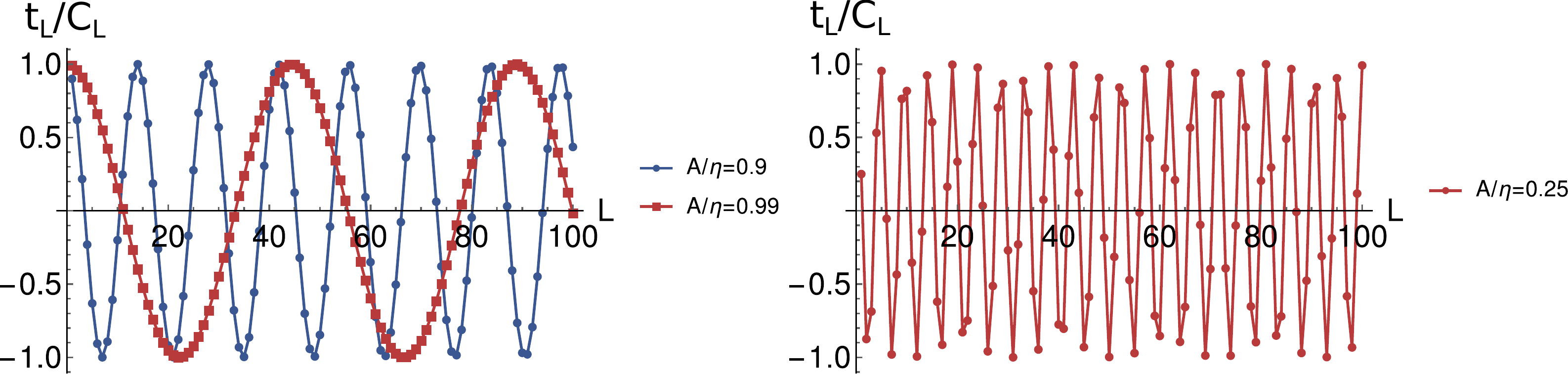}
\par\end{centering}
\caption{$t_{L}/C_{L}$ for different $|A/\eta|<1$.\label{fig:AR_over_etaR}}
\end{figure}

We therefore have that

\begin{equation}
|t_{L}/C_{L}|=|\cos[L\arccos(A/\eta)]|=|T_{L}(A/\eta)|
\end{equation}
where $T_{L}(x)$ is the Chebyshev polynomial of order $L$. It is
very interesting to realize that there is no well-defined limit of
$|t_{L}/C_{L}|$ for large $L$. In fact, even though we always have
$|t_{L}/C_{L}|<1$, its value oscillates with $L$ with a period that
becomes larger the closer $|A/\eta|$ is to $1$ (being infinity at
$|A/\eta|=1$). Examples are shown in Fig.$\,$\ref{fig:AR_over_etaR}.

What about $|A|\geq|\eta|$? In this case, it is useful to use the
following property \citep{Mathematica}:

\begin{equation}
\prod_{\mu=0}^{L-1}\Bigg[x\pm y\cos\Big(\frac{2\pi L'\mu}{L}+\phi\Big)\Bigg]=\frac{1}{2^{L}}\Bigg[\Bigg(x+\sqrt{x^{2}-y^{2}}\Bigg)^{L}+\Bigg(x-\sqrt{x^{2}-y^{2}}\Bigg)^{L}-2(\mp y)^{L}\cos(L\phi)\Bigg].
\end{equation}
This may again be applied to the term $\sum_{\bm{\sigma}\in S_{L}}{\rm sgn}(\sigma)\prod_{\mu=0}^{L-1}e^{i2\pi\tau_{c}\mu\sigma_{\mu}}\eta c_{\mu,-k}(\frac{A}{\eta}+c_{\sigma_{\mu},\phi})$
in Eq.$\,$\eqref{eq:detM_2}, by identifying $x=A/\eta$ and $y=1$.
In this case we obtain 

\begin{equation}
|C_{L}/t_{L}|=\frac{2}{\Bigg(|\frac{A}{\eta}|+\sqrt{(\frac{A}{\eta})^{2}-1}\Bigg)^{L}+\Bigg(|\frac{A}{\eta}|-\sqrt{(\frac{A}{\eta})^{2}-1}\Bigg)^{L}}.
\end{equation}
Note that with $|A/\eta|>1$ this gives an exponential decay for $|C_{L}/t_{L}|$.
This is just what we expect in the extended phase, where the dominant
coupling is $t_{L}$, regarding that it also dominates over $V_{L}$.
At large $L$ we have $|C_{L}/t_{L}|\sim e^{-L/\xi_{c}}$, and the
decay length $\xi_{{\rm EC}}$ is

\begin{equation}
\begin{aligned}\xi_{{\rm EC}}=\frac{1}{\log\Bigg[|A/\eta|+\sqrt{(A/\eta)^{2}-1}\Bigg]}\end{aligned}
,\label{eq:corr_len_ext_crit}
\end{equation}
which is finite for any $|A/\eta|>1$ and of course diverges for $|A/\eta|=1$
(transition between extended and critical phase).

With identical calculations, we can work out the expressions for $|C_{L}/V_{L}|$
just by replacing $A$ with $B$ everywhere above. This yields

\begin{equation}
\begin{aligned}|V_{L}/C_{L}|= & \begin{cases}
|T_{L}(B/\eta)| & ,\textrm{ }|B/\eta|<1\\
\frac{1}{2}\Bigg[\Bigg(|\frac{B}{\eta}|+\sqrt{(\frac{B}{\eta})^{2}-1}\Bigg)^{L}+\Bigg(|\frac{B}{\eta}|-\sqrt{(\frac{B}{\eta})^{2}-1}\Bigg)^{L}\Bigg] & ,\textrm{ }|B/\eta|\geq1
\end{cases}\end{aligned}
.
\end{equation}
For $|B/\eta|\geq1$ we get the following correlation length characterizing
the localized-critical transition:

\begin{equation}
\begin{aligned}\xi_{{\rm LC}}=\frac{1}{\log\Bigg[|B/\eta|+\sqrt{(B/\eta)^{2}-1}\Bigg]}\end{aligned}
.\label{eq:corr_len_loc_crit}
\end{equation}

The ratio $|t_{L}/V_{L}|$ may then be obtained through the previous
expressions. For $|V_{L}/C_{L}|,|t_{L}/C_{L}|>1$ (outside the critical
phase), we have

\begin{equation}
\begin{aligned}|t_{L}/V_{L}|= & \frac{\Bigg(|\frac{A}{\eta}|+\sqrt{(\frac{A}{\eta})^{2}-1}\Bigg)^{L}+\Bigg(|\frac{A}{\eta}|-\sqrt{(\frac{A}{\eta})^{2}-1}\Bigg)^{L}}{\Bigg(|\frac{B}{\eta}|+\sqrt{(\frac{B}{\eta})^{2}-1}\Bigg)^{L}+\Bigg(|\frac{B}{\eta}|-\sqrt{(\frac{B}{\eta})^{2}-1}\Bigg)^{L}}\end{aligned}
.
\end{equation}

We have seen in Ref.~\citep{GoncalvesRG2022} that the correlation
or localization length can be inferred by the scaling of the ratio
$|t_{L}/V_{L}|$ with $L$. We first assume that $|A|<|B|$ (localized
phase). Taking the large $L$ limit we obtain

\begin{equation}
\begin{aligned}\begin{aligned}|t_{L}/V_{L}| & =e^{-L/\xi_{{\rm LE}}},\textrm{ \textrm{with }\ensuremath{\xi_{{\rm LE}}=1/}}\log\left[\frac{|\frac{B}{\eta}|+\sqrt{(\frac{B}{\eta})^{2}-1}}{|\frac{A}{\eta}|+\sqrt{(\frac{A}{\eta})^{2}-1}}\right]\end{aligned}
\end{aligned}
,
\end{equation}
where $\xi_{{\rm LE}}$ is the localization length. We can also compute
the correlation length in the extended phase for $|A|>|B|$. This
can be done just by interchanging $A$ and $B$ in the expression
above, that is:

\begin{equation}
\begin{aligned}\begin{aligned}|V_{L}/t_{L}| & =e^{-L/\xi_{{\rm EL}}},\textrm{ \textrm{with }\ensuremath{\xi_{{\rm EL}}}=1/\ensuremath{\log\left[\frac{|\frac{A}{\eta}|+\sqrt{\left(\frac{A}{\eta}\right)^{2}-1}}{|\frac{B}{\eta}|+\sqrt{\left(\frac{B}{\eta}\right)^{2}-1}}\right]}}\end{aligned}
\end{aligned}
,
\end{equation}
where $\xi_{{\rm EL}}$ is the correlation length.

With all the ratios computed, we can summarize the different phases
as in the main text, with the phase diagram being entirely analytically
determined and only depending on $t_{1}\equiv A(V,\eta,q)$, $V_{1}\equiv B(V,\eta,p)$,
$C_{1}\equiv\eta$.

We will finish this section by cross-checking our correlation length
results for the Aubry-André model. To get exactly the Hamiltonian
in the original paper \citep{AubryAndre}, we need to make $t\rightarrow te^{p}$
and $V\rightarrow Ve^{q}/2$ and then take the large $p$ limit, with
$p=q$. After doing so, we obtain

\begin{equation}
\begin{aligned}\frac{A}{\eta}=-\cosh p+\frac{Ve^{p}}{2\eta}\sinh p\approx\frac{Ve^{2p}}{4\eta}\\
\frac{B}{\eta}=-\cosh p+\frac{te^{p}}{\eta}\sinh p\approx\frac{te^{2p}}{2\eta}
\end{aligned}
.
\end{equation}
This implies that

\begin{equation}
\begin{aligned}\xi_{{\rm LE}}=\frac{1}{\log\Big(\frac{V}{2t}\Big)}\\
\xi_{{\rm EL}}=\frac{1}{\log\Big(\frac{2t}{V}\Big)}
\end{aligned}
,
\end{equation}
which are exactly the correlation lengths previously derived in \citep{AubryAndre}.

\section{Multifractal analysis}

\begin{figure}[h]
\begin{centering}
\includegraphics[width=0.6\columnwidth]{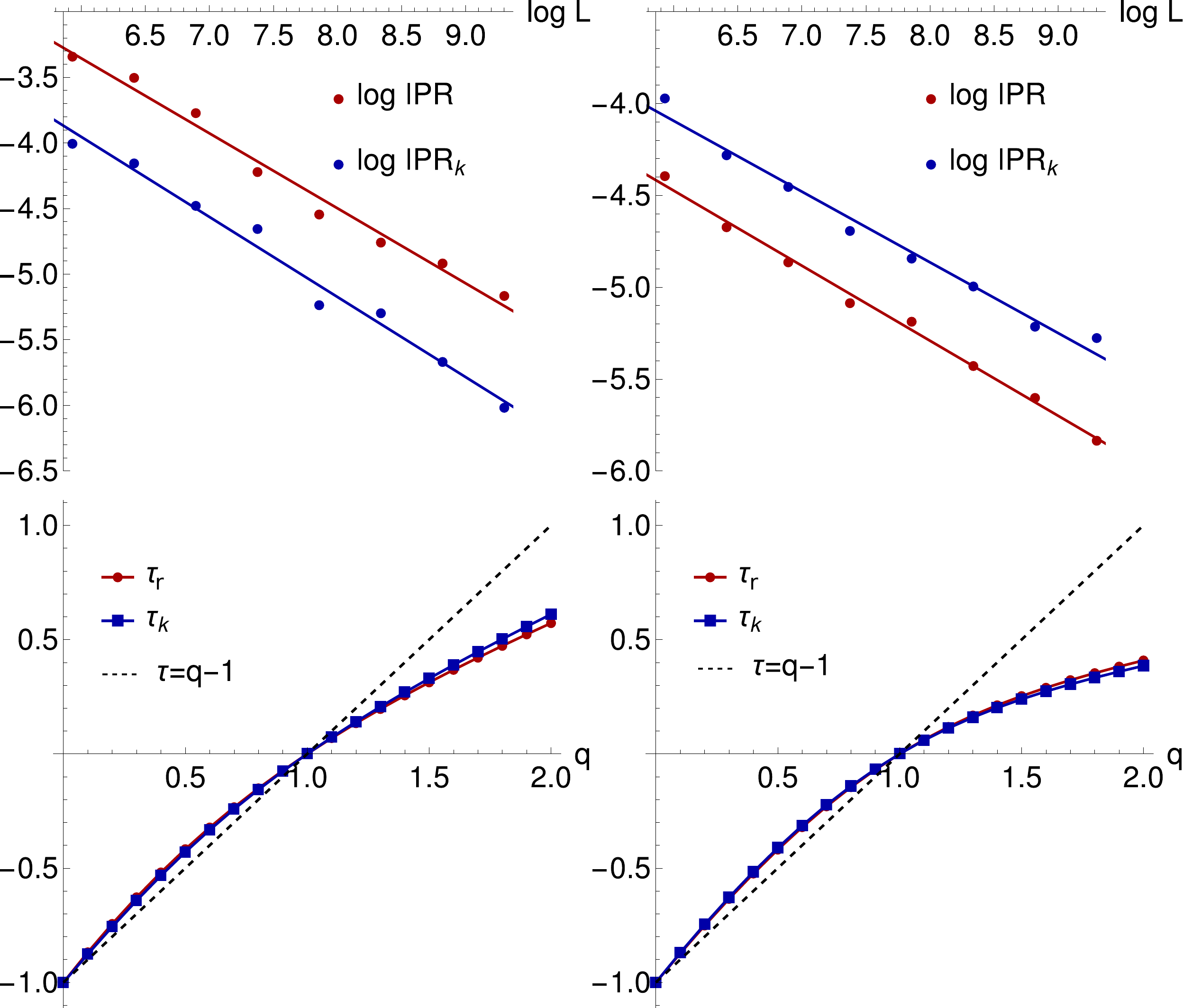}
\par\end{centering}
\caption{(a,b) Scaling of ${\rm IPR}\equiv{\rm IPR}(q=2)$ and ${\rm IPR}_{k}\equiv{\rm IPR}_{k}(q=2)$
with $L$ (top) and multifractal exponents $\tau_{r}(q)$ and $\tau_{k}(q)$
defined in Eq.$\,$\eqref{eq:iprq} (bottom), for a point in the black
path of Fig.$\,$3(c) of the main
text with $V=1.04$ (left) and a point in the cyan path of the same
figure, with $V=0.8$. The results were averaged over a number of
random choices of $\phi$ and $k$ in the interval $[15-750]$, for
system sizes corresponding to the Fibonnacci numbers in the interval
$L\in[377-10946]$ (the largest number of configurations was used
for the smaller sizes). The multifractal exponents were extracting
by fitting (shown in the top panel for $q=2$). The dashed line in
the bottom panel shows the expected behaviour of $\tau_{r}$ ($\tau_{k}$)
if the wave function was fully delocalized in real-space (momentum-space).
\label{fig:multifractal_analysis}}
\end{figure}

In this section, we carry out a multifractal analysis for some points
within the critical phase. To do so, we compute the real- and momentum-space
generalized inverse participation ratios, respectively ${\rm IPR}(q)$
and ${\rm IPR}_{k}(q)$. For an eigenstate $\ket{\psi(E)}=\sum_{n}\psi_{n}(E)\ket n$,
where $\{\ket n\}$ is a basis localized at each site, these quantities
are defined as,

\begin{equation}
\begin{aligned}{\rm IPR}(q)=\Bigl(\sum_{n}|\psi_{n}|^{2}\Bigr)^{-1}\sum_{n}|\psi_{n}|^{2q}\propto L^{-\tau_{r}(q)}\\
{\rm IPR}_{k}(q)=\Bigl(\sum_{m}|\psi_{m}^{k}|^{2}\Bigr)^{-1}\sum_{m}|\psi_{m}^{k}|^{2q}\propto L^{-\tau_{k}(q)}
\end{aligned}
\label{eq:iprq}
\end{equation}
where $\psi_{m}^{k}=L^{-1}\sum_{n}e^{2\pi imm}\psi_{n}$. The size
dependence is characterized by $q$-dependent exponents, $\tau_{r}$
and $\tau_{k}$, defined in terms of the generalized fractal dimensions,
$D_{r}(q)$ and $D_{k}(q)$, as $\tau_{r/k}(q)=D_{r/k}(q)(q-1)$.
Fully extended (localized) states are characterized by $D_{k}(q)=0$
($D_{r}(q)=0$) for $q>0$, and $D_{r}(q)=d$ ($D_{k}(q)=d$). In
these cases $D_{k}(q)$ is constant, and the system is a single-fractal.
Multifractals correspond to cases where $D_{r}(q)$ or $D_{k}(q)$
depends on $q$ \citep{Janssen}. In Fig.$\,$\eqref{fig:multifractal_analysis}
we show that within the critical phase, the generalized fractal dimensions
depend on $q$ as expected in a phase with multifractal properties,
which can be inferred from the non-linear behaviour of $\tau_{r}$
and $\tau_{k}$ with $q$. The wave function is therefore a multifractal
in both real- and momentum-space.

\end{document}